\documentclass[a4paper,11pt]{article}
\usepackage{jheppub} 
\usepackage[section]{placeins}
\usepackage[T1]{fontenc} 
\usepackage{graphicx}
\usepackage{amsmath}
\usepackage{bm}
\usepackage{subfigure}
\usepackage{amssymb}
\usepackage{mathrsfs}
\usepackage{braket}
\usepackage{multirow,booktabs}
\usepackage{comment}

\newcommand{\comments}[1]{}

\usepackage{slashed}
\usepackage{epsfig,psfrag,rotating,soul}
\usepackage{rotfloat}
\usepackage{empheq}  

\allowdisplaybreaks

\title{\boldmath  Study of anomalous charged triple gauge couplings with $W^-W^+$ process at $e^-e^+$ collider}

\author[a,b]{Amir Subba }

\author[c]{, Ritesh K Singh }

\affiliation[a]{ Wilczek Quantum Center, Shanghai Institute for Advanced Studies, Shanghai 201315, China }  

\affiliation[b]{
University of Science and Technology of China, Hefei 230026, China }

\affiliation[c]{Department of Physical Sciences, Indian Institute of Science Education and Research Kolkata, Mohanpur, 741246, India}

\emailAdd{amirsubba@ustc.edu.cn}

\emailAdd{ritesh.singh@iiserkol.ac.in}  

\abstract{
We study anomalous charged triple gauge couplings (aTGC) in $e^-e^+\to W^-W^+$ at $\sqrt{s}=250$~GeV with longitudinally polarized beams, parameterized by the five dimension-six operators $\mathcal{O}_{WWW}$, $\mathcal{O}_{W}$, $\mathcal{O}_{B}$,
$\mathcal{O}_{\widetilde{W}WW}$ and $\mathcal{O}_{\widetilde{W}}$ of the HISZ basis. All three decay channels, $\ell^\pm jj\slashed{E}$, $4j$ and $\ell^-\ell^+\slashed{E}$, are simulated explicitly at a  leading-order including all diagrams that contribute to each final state and hence the interference of $W^-W^+$ with non-resonant amplitudes. The semi-leptonic signal is selected with cuts on $m_{j_1j_2}$ and $m_T^W$. The flavor identity of the two hardest light jets are classified as \emph{up}-type and \emph{down}-type jets with a dedicated boosted decision trees (BDT). In the $4j$ channel a multi-class BDT separates $W^-W^+$ from $ZZ$ and QCD production and the $W^\pm$ candidates are assigned using jet charge. In
the dilepton channel $W^-W^+$ events are selected with cuts on $p_T^{\ell^-}$
and $m_{\ell^-\ell^+}$, and the two neutrino momenta are reconstructed analytically from the on-shell and beam-energy constraints. The cross section and 80 polarization and spin-correlation asymmetries in four disjoint bins of $\cos\theta_{W^-}$ are combined for both beam-polarization configurations through a covariance-based $\chi^2$. The two set of beam polarization offers an orthogonal cuts on parameter space leading to tighter bounds.  The semi-leptonic and full-hadronic final states provides a competitive bounds on aTGCs, while full-leptonic offers the least sensitivity. Marginalized limits on aTGCs via. Markov Chain Monte Carlo analysis for integrated luminosities for different values of integrated luminosities and systematic benchmarks are reported. The projected baseline ILC sensitivity yields stronger constraints on all CP-even aTGC coefficients and $c_{\widetilde{W}}$ than current experimental bounds, while providing competitive sensitivity with respect to HL-LHC projections. Further, with conservative systematic errors, the marginalized limits improves only around $10\%$ from increasing luminosity from $0.5$ ab$^{-1}$ to $8$ ab$^{-1}$, implying the need to control systematics at upcoming colliders. 
}

\begin{document}

\maketitle

\section{Introduction}
\label{sec:intro}

The discovery of weak intermediate vector bosons, $W^\pm$ by the UA1~\cite{UA1:1983crd} and UA2~\cite{UA2:1983tsx} collaboration and $Z$ boson~\cite{UA1:1983mne,UA2:1983mlz} by the same collaboration at CERN supports the $SU(2)_L \times U(1)_Y$ gauge structure of the Standard Model~(SM) of particle physics. The final validation of the SM was done with the discovery of Higgs boson by CMS~\cite{CMS:2012qbp} and ATLAS~\cite{ATLAS:2012yve} collaborations at LHC. Nevertheless, there exist a loophole within the SM that questions the completeness of the SM. The stability of the Higgs mass under radiative corrections, neutrino oscillation and dark matter are some of boiling questions which SM fails to answer. After the discovery of Higgs boson, the precision measurement of the parameters in the Higgs sector could potentially signal the presence of new physics if there is any. Also, directly related to the Higgs sector is the implications of electroweak gauge structure of the SM viz., triple gauge boson couplings (TGC) and quartic gauge couplings (QGC) among the charged and neutral bosons. Thus, measurement of TGC and QGC could directly shed light on gauge structure and the electroweak symmetry breaking (EWSB)~\cite{Higgs:1964ia,Higgs:1964pj,Englert:1964et} of the SM.

Experiments at LHC like CMS~\cite{CMS:2013ant,CMS:2013ryd,CMS:2019efc,CMS:2021icx} and ATLAS~\cite{ATLAS:2012bpb,ATLAS:2012upi,ATLAS:2013way,ATLAS:2016bkj,ATLAS:2016qjc,ATLAS:2021jgw} had extensively studied TGC, viz., $W^-W^+V , V \in \{\gamma,Z\}$ couplings in various di-boson process with semi-leptonic and full-leptonic final decayed states. On the theoretical side, various studies are available that employed different techniques to probe anomalous TGC (aTGC). For instance, Refs.~\cite{Rahaman:2019lab,Rahaman:2019mnz} utilize the polarizations of $W$ boson to constrain aTGC and in study~\cite{Subba:2022czw,Subba:2023rpm,Cheng:2025tym}, additional spin correlations parameters of $WW$ di-boson are incorporated to effectively constrain aTGC. Optimal observable technique (OOT) based limits on aTGC are provided in Ref.~\cite{Diehl:1993br,Jahedi:2024wnw}. Modern machine learning (ML) based techniques has also been used in order to probe aTGC in di-boson process~\cite{Subba:2023rpm,Chai:2024zyl}. The quantum tomography~\cite{Fabbrichesi:2023jep,Aoude:2023hxv} based study of anomalous $W^-W^+\gamma/Z$ has also been done.

The search for the deviation in the $W^-W^+V$ are usually performed in two different framework. One follows the construction of most general Lagrangian involving two charge boson and one neutral boson from which required vertex term are read out. This approach is usually known as LEP framework developed in 1980 assuming $U(1)_{\mathrm{EM}}$ and Lorentz invariance, and the Lagrangian is given in terms of $14$ anomalous parameters as~\cite{Hagiwara:1986vm}
	\begin{align}
		\label{eqn:eff}
		\mathcal{L}_{WWV} &= ig_{WWV}\left[g_1^V(W^+_{\mu \nu}W^{-\mu} - W^{+\mu}W^-_{\mu \nu})V^\nu+  k_V W^+_\mu W^-_\nu V^{\mu \nu} +\frac{\lambda_V}{m_W^2}W_\mu^{\nu+}W_\nu^{-\rho}V_{\rho}^{\mu} \right.\nonumber\\&+\left.ig_4^VW_\mu^+W_\nu^-(\partial^\mu V^\nu+\partial^\nu V^\mu) - ig_5^V\epsilon^{\mu \nu \rho \sigma}(W_\mu^+ \partial_\rho W_\nu^- - \partial_\rho W_\mu^+W_\nu^-)V_\sigma  
		\right.\nonumber\\&+\left.\tilde{k}_VW_\mu^+W_\nu^-\tilde{V}^{\mu \nu} + \frac{\tilde{\lambda}_V}{m_W^2}W_\mu^{\nu+}W_\nu^{-\rho}\tilde{V}_\rho^{\mu}\right],
	\end{align}
where $W^{\pm}_{\mu\nu} = \partial_\mu W^{\pm}_\nu-\partial_\nu W^\pm_\mu$, $V_{\mu\nu} = \partial_\mu V_\nu-\partial_\nu V_\mu$, $g_{WW\gamma} = -e$ and $g_{WWZ} = -e\cot\theta_W$, where $e$ and $\theta_W$ are the proton charge and weak mixing angle respectively. The dual field is defined as $\tilde{V}^{\mu\nu} = 1/2\epsilon^{\mu\nu\rho\sigma}V_{\rho\sigma}$, with Levi-Civita tensor $\epsilon^{\mu\nu\rho\sigma}$ follows a standard convention, $\epsilon^{0123}=1$. Within the SM, the value of the couplings are $g_1^Z=g_1^\gamma = k_Z = k_\gamma = 1$ and all others are zero. The couplings $g_1^V,k^V$, and $\lambda^V$ are CP-even, while $g_4^V$ is odd in C and P-even, $\widetilde{k}^V$ and $\widetilde{\lambda}^V$ are C-even and P-odd and the last coupling $g_5^V$ is C and P-odd which make it CP-even. The search for anomalous signature based on parameterization of Eq.~(\ref{eqn:eff}) has some subtle issues as was pointed out in Ref.~\cite{Degrande:2012wf}; the theory can be expanded to infinitum by adding derivative normalized by mass of $W$ bosons and those terms are not suppressed at energies above $W$ mass, unless the anomalous couplings are very small.

On the absence of concrete evidence for new particles at the LHC, one is generally lead to assume that, new physics (NP) if any, must be too heavy to be observed in the current energy scale. However, even if the NP is too heavy, one can work with the residual effect of such heavy physics at the electroweak scale through some parameters, usually known as Wilson's coefficient (WC). These WCs are associated with higher dimensional~$(d > 4)$ operators constructed out of the SM states. The SM effective field theory (SMEFT) provides a systematic and model-independent framework to parameterize the low-energy effects of heavy new physics beyond the SM in terms of higher-dimensional, $SU(3)_C \times SU(2)_L \times U(1)_Y$ gauge-invariant operators constructed solely from SM fields~\cite{Buchmuller:1985jz,Grzadkowski:2010es}. In this framework, the effective Lagrangian is expressed as an expansion in powers of the cutoff scale $\Lambda$, which denotes the characteristic energy scale of new physics~\cite{Buchmuller:1985jz},
	\begin{equation}
		\mathcal{L}_{\mathrm{EFT}} = \mathcal{L}_{\mathrm{SM}} + \frac{1}{\Lambda^{d-4}}\sum_i c_i^{(d)}\mathcal{O}_i^{(d)},
		\label{eq:SMEFT_Lagrangian}
	\end{equation}
where $\mathcal{O}_i^{(d)}$ are the dimension-$d$ operators constructed from SM fields, and $c_i^{(d)}$ are the corresponding WCs encoding the effects of the ultraviolet (UV) theory. Assuming baryon and lepton number conservation, the lowest-order nontrivial corrections to the SM arise from dimension-6 operators, as dimension-5 terms induce Majorana neutrino masses and violate lepton number~\cite{Weinberg:1979sa}. Given that $\Lambda$ is typically expected to lie in the TeV range, higher-order contributions ($d>6$) can be safely neglected, and the EFT expansion can be truncated at $\mathcal{O}(1/\Lambda^2)$.
	
In this work, we restrict ourselves to the subset of dimension-6 operators relevant for anomalous triple gauge-boson interactions, following the Hagiwara-Ishihara-Szalapski-Zeppenfeld (HISZ) basis~\cite{Hagiwara:1993ck}. The effective Lagrangian with pure aTGC in the HISZ basis can be written in terms of five dimension-6 operators 
	\begin{align}
		\label{eq:HISZ_Lagrangian}
		\mathscr{O}_{WWW} &=  \mathrm{Tr}\!\left[ W_{\nu\rho} W^{\mu\nu} W_\mu^{\ \rho} \right],\nonumber\\
		\mathscr{O}_W &=  (D_\mu \Phi)^\dagger W^{\mu\nu} (D_\nu \Phi),\nonumber\\
		\mathscr{O}_B &=(D_\mu \Phi)^\dagger B^{\mu\nu} (D_\nu \Phi), \nonumber\\ 
        \mathscr{O}_{\widetilde{W}WW} &= \mathrm{Tr}\!\left[ \widetilde{W}_{\mu\nu} W^{\nu\rho} W^\mu_{\ \rho} \right],\nonumber\\
		\mathscr{O}_{\widetilde{W}} &= (D_\mu \Phi)^\dagger \widetilde{W}^{\mu\nu} (D_\nu \Phi),
	\end{align}
	where $\Phi$ is the Higgs doublet, and the covariant derivative and field-strength tensors are defined as
	\begin{align}
		D_\mu &= \partial_\mu + \frac{i}{2} g \tau^i W_\mu^i + \frac{i}{2} g' B_\mu ,\nonumber\\
		W_{\mu\nu} &= \frac{i}{2} g \tau^i (\partial_\mu W_\nu^i - \partial_\nu W_\mu^i + g \epsilon_{ijk} W_\mu^j W_\nu^k) ,\nonumber\\
		B_{\mu\nu} &= \frac{i}{2} g' (\partial_\mu B_\nu - \partial_\nu B_\mu) .
	\end{align}
	Here, $g$ and $g'$ denote the $SU(2)_L$ and $U(1)_Y$ gauge couplings, respectively. Among the operators in Eq.~(\ref{eq:HISZ_Lagrangian}), the first three are CP-even, while the last two are CP-odd. These operators encapsulate the leading-order deviations from the SM gauge-boson self-interactions in a model-independent manner. On matching the WCs with the anomalous parameters in Eq.~\eqref{eqn:eff}, one obtain the following relation between the two theory as
	\begin{align}
		\label{eq:relation}
		&\Delta g_1^Z = \frac{c_Wm_Z^2}{2\Lambda^2},\nonumber\\
        &\Delta \kappa_\gamma = \frac{(c_W+c_B)m_W^2}{2\Lambda^2},\nonumber\\
        &\Delta \kappa_Z = \frac{(c_W-c_B\tan^\theta_W)m_W^2}{2\Lambda^2},\nonumber\\
        &\lambda_\gamma = \lambda_Z = \frac{c_{WWW}3g^2m_W^2}{2\Lambda^2},\nonumber\\
        &g_4^V=g_5^V=0,\nonumber\\
        &\widetilde{\kappa}_\gamma = \frac{c_{\widetilde{W}}m_W^2}{2\Lambda^2},\nonumber\\
        &\widetilde{\kappa}_Z = -\frac{c_{\widetilde{W}}\tan^2\theta_Wm_W^2}{2\Lambda^2},\nonumber\\
        &\widetilde{\lambda}_\gamma = \widetilde{\lambda}_Z = \frac{c_{\widetilde{W}WW}3g^2m_W^2}{2\Lambda^2}.
	\end{align}
	The presence of anomalous couplings at the vertex would modify both the angular and kinematic distributions of final decayed particles from the SM values. The effects of anomalous $W^-W^+\gamma/Z$ couplings on various kinematic and spin-related observables have been theoretically studied in several works~\cite{Buchalla:2013wpa, Choudhury:1999fz, Hagiwara:1992eh, Subba:2022czw, Subba:2023rpm, Rahaman:2019mnz, Bian:2015zha, Bian:2016umx, Tizchang:2020tqs, Rahaman:2019lab}.

	Beyond the triple gauge operators considered in this analysis, there exist additional dimension-6 operators in the SMEFT that contribute to the $e^+e^- \to W^+W^-$ process at the same order in the EFT expansion. In particular, operators that lead to four-point contact interactions ($ff^\prime V_1V_2$) become relevant in $W^+W^-$ production~\cite{Falkowski:2014tna}. In the present work, we focus on the pure bosonic sector and therefore do not include any fermionic operators.

	Most experimental searches for deviations from the SM in $W^-W^+$ production have concentrated on the semi-leptonic channel, which delivers a favorable compromise between reconstruction fidelity and event yield. Nevertheless, the dominant hadronic branching fraction of the $W^\pm$ boson implies that the fully hadronic final state contains a large statistical sample and therefore constitutes a valuable, and in many respects complementary, probe of physics beyond the SM. In this work we perform a comparative study of the $e^-e^+\to W^-W^+$ process in all three channels viz. semi-leptonic ($\ell^\pm jj\slashed{E}$), fully hadronic ($4j$) and fully leptonic ($\ell^+\ell^-\slashed{E}$) final states at a polarized $e^-e^+$ collider, with the aim of constraining anomalous $WW\gamma$ and $WWZ$ couplings in a model-independent SMEFT framework. In particular, we focus on the proposed International Linear Collider (ILC)~\cite{ILC:2007oiw}, which is designed to operate with polarized $e^-e^+$ beams, with electron (positron) polarization degrees of about $80\%$ ($30\%$) along with runs with flip polarization. The machine targets an instantaneous luminosity of $\mathcal{L} \simeq 0.75\times 10^{34}\,\mathrm{cm}^{-2}\,\mathrm{s}^{-1}$~\cite{Behnke:2013xla}, corresponding to an integrated luminosity of roughly $\sim 500\,\mathrm{fb}^{-1}$ over the initial years of running at $\sqrt{s}=250$ GeV. Alongside the ILC, several other proposed $e^-e^+$ facilities such as FCC-ee~\cite{FCC:2018evy}, CEPC~\cite{CEPCStudyGroup:2018rmc}, and CLIC~\cite{CLIC:2016zwp} aim to explore complementary energy regimes with differing center-of-mass energies and luminosity goals.
	
\begin{figure*}[!htb]
		\centering
		\includegraphics[width=0.32\textwidth, height=4.5cm]{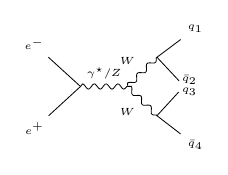}
		\includegraphics[width=0.32\textwidth, height=4.5cm]{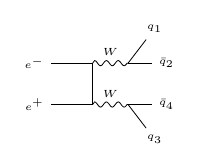}
		\includegraphics[width=0.32\textwidth, height=4.5cm]{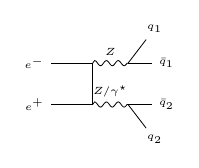}
		\includegraphics[width=0.32\textwidth, height=4.5cm]{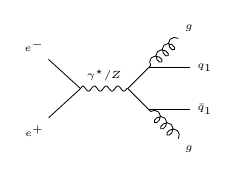}
		\includegraphics[width=0.32\textwidth, height=4.5cm]{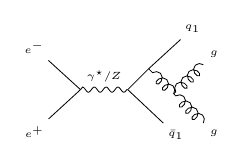}
		\includegraphics[width=0.32\textwidth, height=4.5cm]{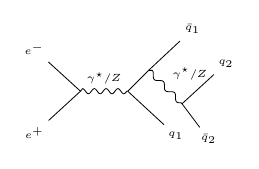}
		\caption{Schematic Feynman diagrams at lowest order for $4j$ events in final state at $e^-e^+$ collider. The top row represents the one/two-resonant $ZZ/\gamma$, and $W^-W^+$ amplitudes with coupling order of $\alpha_{\mathrm{EW}}^2$. The bottom row amplitudes represents gluons in the final state of order $\alpha_{\mathrm{EW}}\alpha_S$, and the zero-resonant amplitudes with four quarks in final states with coupling order of $\alpha_{\mathrm{EW}}^2$. }
		\label{fig:fm_4jet}
	\end{figure*}

	For the four-jet final state the on-shell $W^-W^+$ contribution is embedded within a larger, gauge-invariant set of amplitudes. At leading electroweak order the irreducible backgrounds comprise $ZZ$, $Z\gamma^*$ and other non-resonant continuum topologies contributing at $\mathcal{O}(\alpha_{\mathrm{EW}}^2)$. In addition, reducible QCD-induced topologies with final-state gluons enter at mixed orders such as $\mathcal{O}(\alpha_{\mathrm{EW}}\alpha_S)$ and higher, and must be included for a realistic estimate of the $4j$ final-state rate. Representative LO diagrams for the four-jet topology are shown in Fig.~\ref{fig:fm_4jet}. In the current article, we have considered all possible LO $4j$ amplitudes. 
	
\begin{figure*}[!htb]
		\centering
		\includegraphics[width=0.32\textwidth, height=4.5cm]{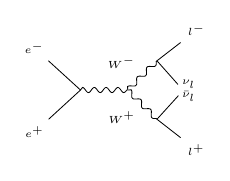}
		\includegraphics[width=0.32\textwidth, height=4.5cm]{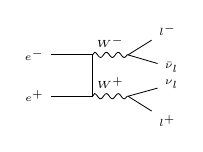}
		\includegraphics[width=0.32\textwidth, height=4.5cm]{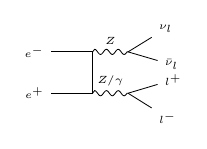}
		\includegraphics[width=0.32\textwidth, height=4.5cm]{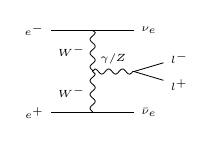}
		\includegraphics[width=0.32\textwidth, height=4.5cm]{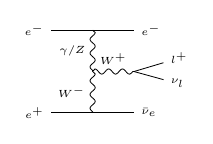}
		\includegraphics[width=0.32\textwidth, height=4.5cm]{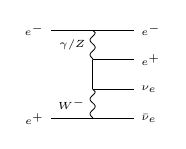}
		\caption{Schematic representation of Feynman diagrams for $e^-e^+ \to l^-l^+ \slashed{E}$ process at the leading order.}
		\label{fig:fm_dilep}
	\end{figure*}

	By contrast, the fully leptonic channel, $e^-e^+\to \ell^+\ell^-\slashed{E}$, has the smallest branching fraction but offers a remarkably clean experimental environment. The schematic LO amplitudes for the dileptonic topology are shown in Fig.~\ref{fig:fm_dilep}. Two well-reconstructed charged leptons and large missing transverse energy provide powerful kinematic handles and strong background suppression. The reconstruction of $W$-boson kinematics, however, is affected by neutrino-related ambiguities, while in the hadronic case, showering and hadronization complicate the interpretation of $WW$ kinematics. Nevertheless, interference between resonant and non-resonant diagrams can enhance sensitivity beyond the statistical advantage of the hadronic mode and partly compensate for the low rate of the leptonic one.  Due to differences in the final states (and hence interferences) they show different sensitivity to anomalous couplings. Consequently, they yield complementary constraints as will be discussed latter.
	
	Although the present analysis is restricted to leading-order (LO) production of $W^-W^+$ at an $e^-e^+$ collider, higher-order quantum corrections from both quantum chromodynamics (QCD) and the electroweak (EW) sector are essential for achieving precision-level predictions. A substantial body of literature has addressed next-to-leading order (NLO) QCD and EW corrections to diboson production at hadron colliders, in particular for the $W^-W^+$ channel at the LHC~\cite{Denner:2023ehn,Frixione:1993yp,Ohnemus:1994ff,Caola:2023wpj}, as well as at $e^-e^+$ colliders~\cite{Denner:2000bj,Denner:2005es}. While inclusive EW corrections typically induce modest shifts in the total cross section, they can lead to sizable distortions in differential observables, especially in the high-energy tails of transverse momentum and invariant mass distributions. The interplay between aTGC and higher-order corrections has also been investigated in vector boson pair production~\cite{Baur:1989qt,Baur:1995uv,Baglio:2017bfe,Chiesa:2018lcs}, demonstrating that NLO effects can be comparable to, and sometimes mimic, deviations induced by aTGC operators. Furthermore, polarization and spin-correlation observables of weak bosons have been studied within the SMEFT framework at NLO QCD accuracy in hadron colliders~\cite{Rahaman:2019lab,Pelliccioli:2026ltl}. A comprehensive analysis of the $W^-W^+$ process at $e^-e^+$ collider, including all decay channels and incorporating full NLO quantum corrections, is deferred to future work.
	
	The remainder of this article is organized as follows. Sec.~\ref{sec:obs} introduces the formalism for describing the $W^+W^-$ polarization states and spin correlations, and outlines the extraction of these observables from joint angular distributions of the final-state jets. Sec.~\ref{sec:collider}, discuss the collider simulation pipeline and the input electroweak parameters. Sec.~\ref{sec:semi_lep} discuss the analysis of semi-leptonic channel of $W^-W^+$ process along with flavor tagging using boosted decision trees (BDT). The two hardest light jets are classified into \emph{up}- or \emph{down}-type categories, as described sub-section~\ref{sec:flav}. Sec.~\ref{sec:fullhad} details the collider simulation and event-selection strategy in full-hadronic final states, including signal--background separation using a BDT. After selecting $W^+W^-$-like signal events, jet-charge based tagging is applied to assign $W^-$ and $W^+$ candidates, discussed in sub-section~\ref{sec:jetcharge}. In sec.~\ref{sec:full-lep}, we discuss the collider analysis of full-leptonic channel along with the reconstruction of two missing neutrinos using $W$ mass constraints and beam energy. The analysis and results on anomalous $WW\gamma$ and $WWZ$ couplings within the SMEFT framework are presented in sec.~\ref{sec:results} We finally conclude in sec.~\ref{sec:concl}.

	\section{Angular distribution}
	\label{sec:obs}
	The spin of a particle is a fundamental property that influences the Lorentz structure of its interactions with other SM particles, and understanding these interactions is essential for probing SM or any beyond SM theories~\cite{Boudjema:2009fz,Aguilar-Saavedra:2017zkn,Aguilar-Saavedra:2015yza}. On the other hand, the spin of a particle also dictates the angular distribution of decayed particles, which can be influenced by the change in the interaction. Thus, the change in couplings due to the presence of new physics may lead to deviation of various angular functions of final state particles from the SM value. Based on the value of spin of a decaying particle, we can quantify these angular functions in terms of polarization parameters. In particular, for a spin-1 particle like $W$ boson, eight independent polarization parameters holds the information of production dynamics which can be obtained by using the angular distribution of the decayed daughter. And in a process where two spin-1 boson~($W^-W^+$) are produced, the production dynamics are encoded in $16$ polarizations, and $64$ spin correlations, which can be obtained in terms of the polar and azimuth angle of final decayed fermions.

 For $W^-W^+ \to f_1f_1^\prime f_2f_2^\prime$ decay process, the joint angular distribution of the final decayed fermions are written as~\cite{Rahaman:2021fcz}
	\begin{align}
		\label{eqn:jdm}
		\frac{1}{\sigma}\frac{d\sigma}{d\Omega_{f_1}d\Omega_{f_2}} &= \sum_{\lambda^\prime s}\rho_{W^-W^+}\left(\lambda_{W^-},\lambda_{W^-}^\prime,\lambda_{W^+},\lambda_{W^+}^\prime\right)\nonumber\\&\times \Gamma_{W^-}\left(\lambda_{W^-},\lambda_{W^-}^\prime\right)\Gamma_{W^+}\left(\lambda_{W^+},\lambda_{W^+}^\prime\right),			
	\end{align} 
	where the production density matrix $\rho_{W^-W^+}$ can be written in basis space of $9\times 9$ matrices formed by the tensor product $\{\mathbb{I}\otimes \mathbb{I},\mathbb{I}\otimes J_i,J_i \otimes \mathbb{I}, J_i\otimes J_j \}$. Then, the fully correlated  density matrix is given by
	\begin{align}
		\label{eqn:rho}
		\rho_{W^-W^+} &= \frac{1}{9}\mathbb{I}\otimes \mathbb{I} + \sum_{i=1}^8\left(P_i^{(1)}J_i^{(1)}\otimes \mathbb{I}+P_i^{(2)} \mathbb{I}\otimes J_i^{(2)}\right) + \sum_{i,j=1}^8C_{ij}^{(12)}J_i^{(1)}\otimes J_j^{(2)}.
	\end{align}
	Here, $P^\prime$s are eight independent polarization, and $C_{ij}^{(12)}$ are $64$ correlation  parameters. The $J_i^\prime$s matrices are constructed from the three spin-1 operators $S_i, i \in \{1,2,3\}$ and the $J_i^\prime$s matrices have orthonormal properties $\text{Tr}\left[J_iJ_j\right] = \delta_{ij}/2,$ for $i,j > 0$ (3 if $i = j = 1)$. The normalized decay density matrix $\Gamma_{W}$ is given by~\cite{Boudjema:2009fz}
        \begin{equation}
				\label{eq:decaymatrix}
				\Gamma(\lambda,\lambda^\prime) = \begin{bmatrix}
					\frac{1+\delta+(1-3\delta)\cos^2\theta+2\alpha\cos\theta}{4}&\frac{\sin\theta(\alpha+(1-3\delta)\cos\theta)}{2\sqrt{2}}e^{i\phi}&(1-3\delta)\frac{1-\cos^2\theta}{4}e^{i2\phi}\\\\
					\frac{\sin\theta(\alpha+(1-3\delta)\cos\theta)}{2\sqrt{2}}e^{-i\phi}&\delta+(1-3\delta)\frac{\sin^2\theta}{2}&\frac{\sin\theta(\alpha-(1-3\delta)\cos\theta)}{2\sqrt{2}}e^{i\phi}\\\\
					(1-3\delta)\frac{(1-\cos^2\theta)}{4}e^{-i2\phi}&\frac{\sin\theta(\alpha-(1-3\delta)\cos\theta)}{2\sqrt{2}}e^{-i\phi}&\frac{1+\delta+(1-3\delta)\cos^2\theta}{4}2\alpha\cos\theta
				\end{bmatrix},
			\end{equation}
	where $\theta$, $\phi$ are respectively, the polar and azimuth orientation of the final decayed fermion in the rest frame of $W$ boson.
     The parameter $\alpha$ is spin analyzing  power of final fermion. For the decay of spin-1 boson to two fermions $(V \to f_1f_2)$ with decay vertex of the form $\bar{f}_1\gamma^\mu(C_LP_L+C_RP_R)f_2V_\mu$ with real $C_{L,R}$, $\alpha$ in the rest frame of $V$ takes the form~\cite{Boudjema:2009fz}
	\begin{align}
		\alpha = \frac{2(C_R^2-C_L^2)\,G(x_1,x_2)}{(C_R^2+C_L^2)G_2(x_1,x_2)+12C_LC_Rx_1x_2},
	\end{align}
	where $x_i = m_i/m_V$ and $G_i$ are defined as
    \begin{align}
        G_1(x_1,x_2) &=\sqrt{1+(x_1^2-x_2^2)^2-2(x_1^2+x_2^2)},\nonumber\\
        G_2(x_1,x_2) &=2-(x_1^2-x_2^2)^2+(x_1^2+x_2^2).
    \end{align}
For the decay $W \to \ell\nu/j_1j_2$, with $m_i \to 0$ at high energy limit and within the SM we have $C_R= 0 $ leading to $\alpha = -1$. Similarly, the expression for the parameter $\delta$ is obtained from the decay amplitude and is given by~\cite{Boudjema:2009fz}
	\begin{align}
		\delta = \frac{(c_R^2+C_L^2)\left[(x_1^2+x_2^2)-(x_1^2-x_2^2)^2\right]+4C_LC_Rx_1x_2}{(C_R^2+C_L^2)\,G_2(x_1,x_2)+12C_LC_Rx_1x_2},
	\end{align}
	which at the high energy limit goes to zero.

    For the single $W$ boson, the angular distribution of final state fermions in the rest from of decaying $W$ boson is given by Eq.~\eqref{eqn:jdm} as,
\begin{align}
\frac{1}{\sigma}\frac{d\sigma}{d\Omega_f} &= \frac{1}{16\pi}\Bigl[4
+ 6\alpha\sin\theta\cos\phi\,P_1 + 6\alpha\sin\theta\sin\phi\,P_2
+ 6\alpha\cos\theta\,P_3 \nonumber\\&+ (1-3\delta)\Bigl(
3\sin^2\theta\sin 2\phi\,P_4
+ 6\sin\theta\cos\theta\cos\phi\,P_5
+ 6\sin\theta\cos\theta\sin\phi\,P_6
\\&+ 3\sin^2\theta\cos 2\phi\,P_7
+ \sqrt{3}\,(3\cos^2\theta-1)\,P_8
\Bigr)\Bigr],\nonumber
\end{align}
where $d\Omega_f = d\cos\theta\,d\phi$.
    The polarization and spin correlation parameters of Eq.~\eqref{eqn:rho} can be obtained through partial integration of joint angular distribution w.r.t to polar and azimuth angles~\cite{Rahaman:2023pte}. Numerically, these parameters are computed from the asymmetries in the angular functions as,
	\begin{equation}
		A_{ij} = \frac{\sigma(f_i^{(1)}f_j^{(2)} > 0) - \sigma(f_i^{(1)}f_j^{(2)}<0)}{\sigma(f_i^{(1)}f_j^{(2)} > 0) + \sigma(f_i^{(1)}f_j^{(2)}<0)},
	\end{equation}
	where $A_{i0}$ and $A_{0j}, i,j \in {1,8}$, are the polarization asymmetries for $W^-$ and $W^+$, respectively. And $A_{ij}, i,j \in \{1,..,8\}$ are the spin correlation asymmetries. The parameters $f^{(1/2)}_i, i \in \{1..8\}$ are the corresponding correlators or angular functions associated with the two distinct final fermions, which are listed in Table~\ref{tab:corr}. One can obtain the exact relation between the asymmetries with polarization, and spin correlation parameters by doing a partial integration of Eq.~(\ref{eqn:jdm})~(see Ref.~\cite{Rahaman:2021fcz}).

\begin{table}[!htb]
    \centering
    \caption{\label{tab:corr}List of angular functions that are associated with the distribution of final decayed fermions and the corresponding generators.}
    \renewcommand{\arraystretch}{1.3}
    \begin{tabular}{ccc}
        \hline
        Correlators & Functions & Generators \\
        \hline
        $f_0$ & $1$ & $J_0=\mathbb{I}$\\
        $f_1$ & $\sin\theta\cos\phi$ & $J_1 = S_x/2$ \\
        $f_2$ & $\sin\theta\sin\phi$ & $J_2 = S_y/2$ \\
        $f_3$ & $\cos\theta$ & $J_3 = S_z/2$ \\
        $f_4$ & $\sin^2\theta\sin(2\phi)$ & $J_4 = \left(S_1S_2 + S_2S_1\right)/2$\\ 
        $f_5$ & $\sin(2\theta)\cos(\phi)$ & $J_5 = \left(S_1S_3+S_3S_1\right)/2$\\
        $f_6$ & $\sin(2\theta)\sin(\phi)$ & $J_6 = \left(S_2S_3+S_3S_2\right)/2$\\
        $f_7$ & $\sin^2\theta\cos(2\phi)$ & $J_7 = \left(S_1S_1 - S_2S_2\right)/2$\\
        $f_8$ & $3\cos^2(\theta)-1$ & $J_8=\sqrt{3}\left(S_3S_3/2-\mathbb{I}/3\right)$\\
        \hline
    \end{tabular}
\end{table}
 

	The correlators $f_i$'s have distinct transformation properties under CP. For instance, the correlator $f_1$ is CP-even, while $f_2$ is CP-odd. The construction of asymmetries associated with vector polarizations, as well as vector--vector and vector--tensor correlations, requires tagging of the decay products of the $W$ bosons. Since we analyze processes corresponding to the $W^-W^+$ topology, the decay products of the $W$ bosons are identified according to their flavor structure. For hadronic decays, where the final states consist of quark jets, we employ a boosted decision tree~(BDT)--based flavor-tagging algorithm to distinguish jets originating from \emph{up}- and \emph{down}-type quarks. In contrast, for fully leptonic decay channels, we assume ideal lepton identification.
	
	
	In the subsequent sections, we initiate the collider-level analysis, including event selection, flavor tagging and neutrino reconstruction, in all three decay channels of $WW$ di-boson processes.

	\section{Collider Simulation}
	\label{sec:collider}
    In the current work, the collider simulation proceeds as follows.
    The dimension-6 operators listed in Eq.~\eqref{eq:HISZ_Lagrangian} are implemented in \textsc{FeynRules}~\cite{Alloul:2013bka,Christensen:2008py} to obtain \textsc{Universal FeynRules Output}~\cite{Degrande:2011ua,Darme:2023jdn} (UFO) model for generation of Monte Carlo events. The matrix level events are generated using \textsc{ MadGraph5$\_$aMC$@$NLO} (v.7.3)~\cite{Alwall:2011uj,Frederix:2018nkq} (MG5 henceforth) at $\sqrt{s}=250$ GeV. The electroweak input parameters follows that of LEP or $\alpha$-scheme~\cite{Biekotter:2023xle} which uses $\{\alpha_{\mathrm{EW}},G_F,m_Z\}$ as inputs,  where the SM input parameters are defined as,
	\begin{align}
			&e = \sqrt{4\pi \alpha_{\mathrm{EW}}},~ g = \frac{e}{\cos\theta},~g^\prime = \frac{e}{\sin\theta},
			~v = \frac{1}{2^{1/4}\sqrt{G_F}},\nonumber\\ &\sin^2\theta= \frac{1}{2}\left[1-\sqrt{1-\frac{4\pi\alpha_{\mathrm{EW}}}{\sqrt{2}G_Fm_Z^2}}\right],\\&m_W^2 = m_Z^2\cos^2\hat{\theta}\nonumber.
	\end{align}
	The showering of colored partons and hadronization of final state particles are done using \textsc{ Pythia}(v$8.3.06)$~\cite{Bierlich:2022pfr} with Monash 2013 Tune~\cite{Skands:2014pea}. Finally, the detector effects are implemented with \textsc{ Delphes}~\cite{deFavereau:2013fsa} with publicly available International Linear Collider~(ILC) simulation parameters~\cite{Potter:2016pgp} in \textsc{Delphes}. All events are generated at a center-of-mass energy of $250~\text{GeV}$ with the following parton-level kinematic selections:
		\begin{align}
			&p_T^j \ge 20~\mathrm{GeV},\quad p_T^\ell \ge 10~\mathrm{GeV},\quad \mathrm{max}\left(|\eta_{j/\ell}|\right) = 2.5,\nonumber\\&\mathrm{min}\left(\Delta R_{ij}\right) = 0.4,        
		\end{align}
	where $\eta_{j/\ell}$ is the pseudorapidity of jets/leptons and $\Delta R_{ij} = \sqrt{(\eta_{i}-\eta_{j})^2 + (\phi_{i}-\phi_{j})^2}$ is the geometric distance between two particles (jets, leptons). 
	\section{The $e^-e^+ \to \ell^\mp jj\slashed{E}$ process}
	\label{sec:semi_lep}
    \begin{figure*}[!htb]
		\centering
		\includegraphics[width=0.49\textwidth]{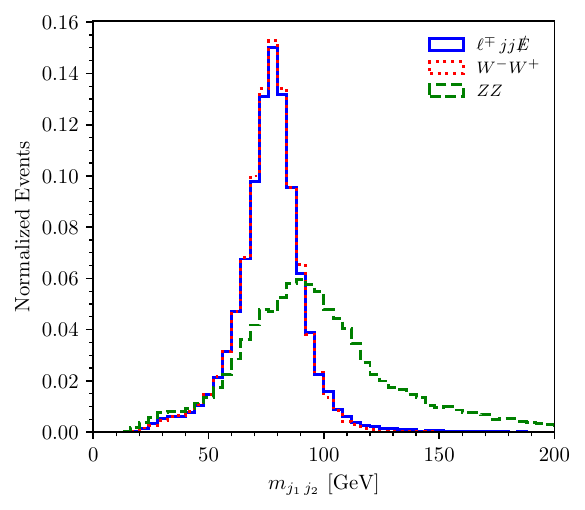}
		\includegraphics[width=0.49\textwidth]{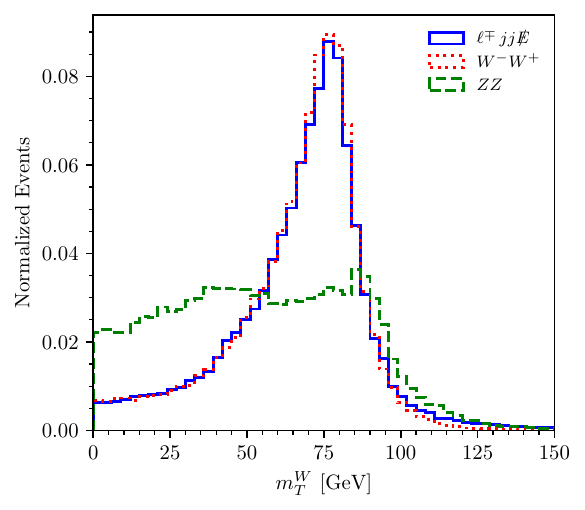}
		\caption{\label{fig:semi_kin} Normalized distribution of invariant mass of two hardest jets and the transverse mass of charged lepton and missing energy. The distribution are obtained for $\ell^\pm jj \slashed{E}$ along with semi-leptonic decay of $W^-W^+$ and $ZZ$ di-boson process at $\sqrt{s}=250$ GeV.}
	\end{figure*}
	We simulate $e^-e^+ \to \ell^\mp jj\slashed{E}$ process following the above discussion. Events are selected with exactly one charged leptons which satisfy the isolation criteria defined as~\cite{deFavereau:2013fsa} $I(\ell) = \sum_{i\neq \ell } p_T^i/p_T^{\ell},$ where $p_T^{\ell}$ is the transverse momentum of the lepton under consideration, and the sum runs over all other particles within a cone of radius $\Delta R$ around the lepton axis satisfying $p_T^i > p_T^\mathrm{min}$. We adopt $p_T^{\mathrm{min}} = 0.5~\mathrm{GeV}$ and $\Delta R = 0.5$ uniformly for both electrons and muons. For electron, we choose $I(e)^{\mathrm{max}} = 0.12$, while for muon it is set to $0.25$. The irreducible background considered is $ZZ \to \ell^-\ell^+ j j$ process in which one of the lepton do not satisfy the isolation criteria and is registered as an missing vector. 

   To reconstruct the $W^-$ boson, we recover the neutrino momentum using the on-shell $W$ mass constraint, $(p_\ell + p_\nu)^2 = m_W^2$. This yields a quadratic equation in the neutrino longitudinal momentum, $p_{\nu z}$, with
two solutions. We choose the solution with the smaller $|p_{\nu z}|$. 

	To enhance the signal purity, we further require $m_{j_1 j_2} \le 90$ GeV and $50 \le m_T^W \le 100$ GeV, where $m_{j_1 j_2}$ is the invariant mass of the two leading jets and $m_T^W$ the transverse mass of the leptonically decaying $W$ boson. After these cuts, the residual 	$ZZ$ contribution constitutes approximately $0.7\%$ of the selected 	sample, corresponding to $\sim 400$ events at an integrated luminosity
	of $\mathcal{L} = 1000$~fb$^{-1}$. For the inclusive 	$\ell^\mp jj \slashed{E}_T$ final state, the overall selection 	efficiency is $32\%$, yielding $\sim 1.5 \times 10^5$ $W^-W^+$ events. 	Owing to this suppression, the $ZZ$ background is neglected in the 	remainder of the analysis.

    In the subsequent section, we discuss the flavor tagging of two hardest jets from the $W$ boson decay using boosted decision trees.

    \subsection{Jet Flavor Tagging}
	\label{sec:flav}
	The endeavor of jet tagging, which involves identifying the origin of jets produced in high-energy collisions, has been a pivotal aspect of particle physics research. Historically, significant attention was directed towards distinguishing quark-initiated jets from gluon-initiated jets, leveraging the inherent differences in their color charge and associated radiation patterns~\cite{Altarelli:1977zs}. The Altarelli-Parisi splitting functions elucidate that gluons, characterized by a higher color factor ($C_A = 3$) compared to quarks ($C_F = 4/3$), tend to radiate more, resulting in jets with a higher constituent multiplicity and broader radiation patterns. This foundational understanding was instrumental in early experimental investigations conducted by different collaborations such as SLAC~\cite{Nilles:1980ys,Derrick:1985du,Petersen:1985hp}, JADE~\cite{JADE:1982ttq,JADE:1983ihf}, TASSO~\cite{TASSO:1989tbj}, AMY~\cite{AMY:1989rdg}, DELPHI~\cite{Marti:1995im}, and OPAL~\cite{OPAL:1993uun,OPAL:1991ssr}.
	
	At lower energy scales, the structural similarities between gluon jets and $b$-jets, particularly in terms of constituent multiplicity due to the extended decay chains of $B$-hadrons~\cite{ALEPH:1995oxo}, posed challenges in differentiation. However, at the Large Hadron Collider (LHC), the higher transverse momenta ($p_T$) have accentuated the distinctions between gluon and $b$-jets, primarily due to the increased color activity inherent in Quantum Chromodynamics (QCD) processes. These color properties had been utilized in ATLAS~\cite{ATLAS:2014vax,ATLAS:2016wzt,ATLAS:2017nma} and CMS~\cite{CMS:2013kfa} for tagging jets initiated by gluons and quarks.
	
	Further, the advent of machine learning (ML) techniques has revolutionized jet tagging methodologies. Initial studies demonstrated that employing multivariate analyses~\cite{Gallicchio:2011xq}, utilizing variables such as charged track multiplicity and $p_T$-weighted linear radial moments, could achieve performance levels comparable to traditional methods. Subsequent research expanded the repertoire of jet substructure observables, incorporating metrics like Les Houches Angularity, width, thrust, and $(p_T^D)^2$, to enhance tagging accuracy~\cite{CMS:2021iwu}. The jet charge variable has also been studied for tagger purposes~\cite{CMS:2017yer}. 
	
	Machine learning has further transformed jet tagging by surpassing traditional feature-based methods through data-driven approaches. Early applications used convolutional neural networks~(CNNs) to process jets as images, enhancing quark/gluon discrimination and boosted object tagging~\cite{ATLAS:2017dfg,Komiske:2016rsd,Szegedy:2014nrf,Barnard:2016qma,Almeida:2015jua,deOliveira:2015xxd,Cogan:2014oua,Andrews:2019faz,Lee:2019cad}. Graph-based models like ParticleNet improved performance by representing jets as graphs with kinematic relationships encoded in edges~\cite{Qu:2019gqs}. Transformer models further advanced tagging via self-attention, capturing long-range correlations among jet constituents~\cite{He:2023cfc,Qu:2019gqs}. Energy Flow Networks (EFNs) introduced a permutation-invariant approach leveraging energy flow polynomials for full kinematic representation~\cite{Komiske:2018cqr}. Unsupervised and weakly supervised methods reduce dependence on labeled data, utilizing Poissonian likelihood-based tagging and weak supervision~\cite{Alvarez:2021zje,Dolan:2023abg}. Likelihood discriminants remain a key traditional tool~\cite{Cornelis:2014ima}, while DeepJet integrates diverse jet features to boost efficiency~\cite{Bols:2020bkb}. Among ML architectures, transformers currently achieve the highest classification accuracy~\cite{Qu:2019gqs}. While deep learning dominates, boosted decision trees (BDT) based algorithm remain crucial for fast, interpretable classification in real-time event selection~\cite{ATLAS:2023dyu}.
	
In the present analysis we do not aim at a complete quark-flavor identification, but specifically at separating \emph{up-type} from \emph{down-type} jets originating from $W$-boson decays. The two leading jets in our events arise from hadronic $W$ decays and, within the SM, are initiated by light quarks ($u,d,c,s$). Knowledge of whether a jet is \emph{up}- or \emph{down}-type is essential for reconstructing polarization and spin-correlation observables that probe anomalous $W^-W^+\gamma/Z$ couplings; without flavor tagging, the vector polarization  asymmetries average out and the sensitivity is reduced~\cite{Subba:2023rpm}.

We perform the \emph{up/down} discrimination with a gradient-boosted decision tree (BDT) implemented via the XGBoost
framework~\cite{Chen_2016}. The input features are built from the constituents of the two hardest jets, reconstructed at detector level with \textsc{Delphes} at $\sqrt{s} = 250$~GeV, and are summarized in Table~\ref{tab:flavtag}.They comprise the multiplicities of the reconstructed photons,
leptons, charged pions and charged kaons in each jet, $N_\gamma$, $N_\ell$, $N_{\pi^\pm}$ and $N_{K^\pm}$. The corresponding four-momenta $p_\gamma$, $p_\ell$, $p_\pi$ and $p_K$ enter with their continuous components normalized by the jet energy, together with the scalar sums of transverse momentum, $p_T^\gamma$, $p_T^\ell$,
$p_T^{\pi^\pm}$ and $p_T^{K^\pm}$. We further include the $p_T$-weighted jet charge defined as~\cite{Field:1977fa,Kang:2023ptt}
\begin{align}
	\label{eq:jet_charge}
	Q_J^\kappa = \frac{1}{p_{TJ}^\kappa} \sum_{i=1}^n Q_i\, p_{Ti}^\kappa,
\end{align}
where $p_{TJ}$ is the transverse momentum of the jet, and $Q_i$ and $p_{Ti}$ denote the electric charge and transverse momentum of its constituents. The exponent $\kappa$ is a phenomenological parameter
controlling the relative weight of soft and hard constituents, typically chosen in the range $0 \le \kappa \le 1$. Finally, the number of displaced mother particles $N_d$ is defined as those with a displacement
$d > 0.3$~mm from the primary vertex. The latter probes displaced tracks and is particularly sensitive to the short-lived $K^0_S$ mesons that accompany $c$-quark-initiated jets.

Truth labeling is performed by matching quarks to jets through the geometric distance $\Delta R_{qj}$. Among the four possible quark--jet pairings, the one with the smallest $\Delta R$ is retained; in case of degeneracy, the harder jet is assigned to the quark.

\begin{table}[!t]
	\centering
	\renewcommand{\arraystretch}{1.5}
	\caption{\label{tab:flavtag} List of observables used as input
		features for the boosted decision tree to classify the two leading
		jets as either \emph{up/down}-type jets. The features are derived
		at $\sqrt{s}=250$~GeV at the detector (\textsc{Delphes}) level.}
	\begin{tabular*}{0.99\textwidth}{@{\extracolsep{\fill}}cc@{}}
		\hline
		Feature & Description  \\
		\hline
		\( N_\gamma \), \( N_\ell \), \( N_{\pi^\pm} \), \( N_{K^\pm} \) & Multiplicity of the jet constituents \\
		\( p_\gamma, p_\ell, p_\pi, p_K \) & Four-momentum  \\
		\( p_T^\gamma, p_T^\ell, p_T^{\pi^\pm}, p_T^{K^\pm} \) & Scalar sum of \( p_T \)\\
		\( Q_J^\kappa \) & Jet charge (\( \kappa \in \{0,1\} \))  \\
		$N_d$ & mother particles with lifetime $d > 0.3$ mm \\
		\hline
	\end{tabular*}
\end{table}
We consider the binary 	task of separating \emph{up}-type from \emph{down}-type quark jets, where the
	prediction for an event $i$ is the logistic probability
	\begin{equation}
		\hat{y}_i = \sigma(z_i) = \frac{1}{1 + e^{-z_i}},\qquad	z_i = \sum_{k=1}^{K} f_k(\mathbf{x}_i),
		\label{eq:bdt-sigmoid}
	\end{equation}
	with $\mathbf{x}_i$ the input feature vector and $f_k$ the $k$-th regression tree. Training minimizes the regularized objective
	\begin{equation}
		\mathcal{L} = \sum_{i=1}^{N}\ell\!\left(y_i, \hat{y}_i\right)+\sum_{k=1}^{K} \Omega(f_k),
		\label{eq:bdt-objective}
	\end{equation}
	where the first term is the binary cross-entropy (log-loss)
	\begin{equation}
		\ell(y, \hat{y}) =-\left[\, y \log \hat{y}+ (1-y) \log(1-\hat{y})\,\right],
		\label{eq:bdt-logloss}
	\end{equation}
	and the second term is the tree-complexity penalty
	\begin{equation}
		\Omega(f) = \gamma\, T + \tfrac{1}{2}\lambda \|w\|_2^2+ \alpha \|w\|_1,
		\label{eq:bdt-regularization}
	\end{equation}
	with $T$ the number of leaves, $w$ the leaf weights, and
	$(\gamma, \lambda, \alpha)$ the corresponding regularization
	strengths.
	
	The hyperparameters are tuned by a randomized search over
	$30$ candidate configurations with $3$-fold cross-validation,
	optimizing the negative log-loss on the validation fold. The best
	configuration found is $\eta = 0.05$, maximum depth of 6,
	sub-sample of 0.8, $\gamma = 0.0$, $\lambda = 3.0$, and $\alpha = 0.1$, with the	number of boosting rounds fixed at $2000$ and early stopping 	activated after $50$ non-improving rounds. To compensate for the
	residual class imbalance, per-event weights 	$w_c = N / N_c$ are applied during training, where $N_c$ is the 	number of training events in class $c \in \{\text{up}, 	\text{down}\}$. Input features are standardized to $[0,1]$ with a 	min--max scaler fitted on the training set and applied identically
	to the validation and test sets.
	
\begin{figure}[!htb]
		\centering
		\includegraphics[width=0.49\textwidth]{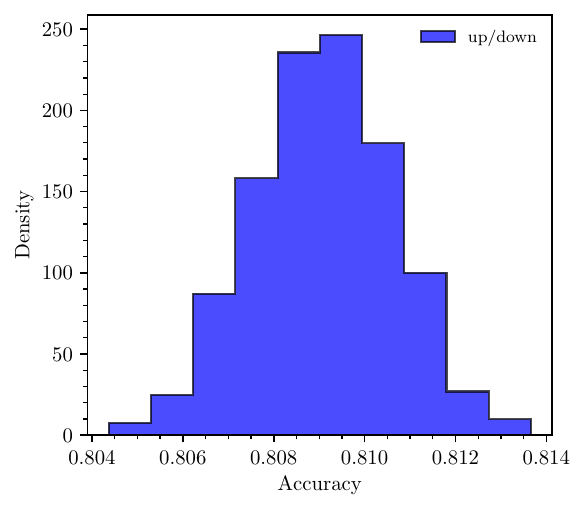}
        \includegraphics[width=0.49\textwidth, height=6.5cm]{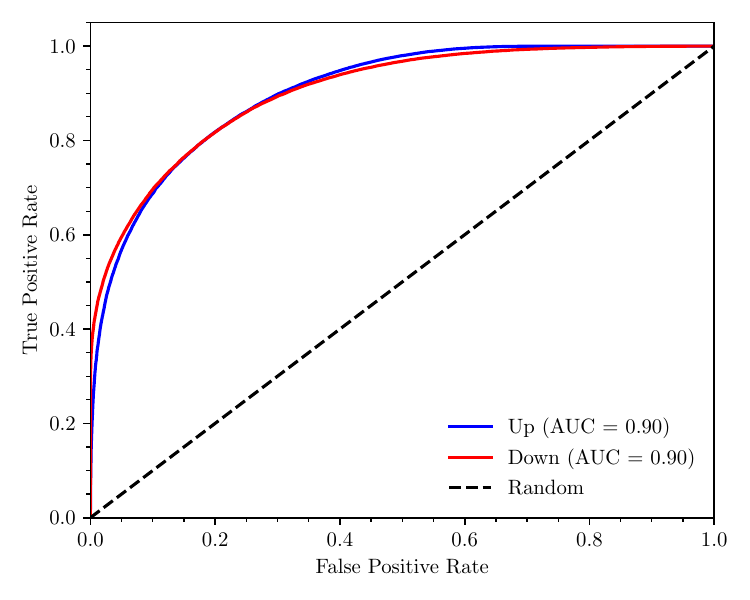}
		\caption{\label{fig:flavacc}Distribution of accuracy (left panel) and ROC curve (right panel) of classifying the \emph{up}/\emph{down}-type quark jets from the decay of $W$ boson using boosted decision tree. The accuracy are obtained with classifier threshold of $0.5$. The training sample are obtained from semi-leptonic decay of $W^-W^+$ process simulated at $\sqrt{s}=250$ GeV $e^-e^+$ collider.}
	\end{figure}

	The model was trained on unpolarized datasets, as these yielded performance comparable to polarized samples~\cite{Subba:2023rpm}. The model reaches an area under the curve (AUC) of $0.9$ of ROC curve. Further, the model classification accuracy was evaluated through iterative sub-sampling of test data, achieving approximately 81\% mean accuracy with a variance of 0.01\%. The distributions of classification accuracy for $W^-$ and $W^+$ jets are shown in Fig.~\ref{fig:flavacc}. The current method of flavor tagging is independent of different choice of parton showering and hadronization model~\cite{Subba:2022czw}.
	
\section{The $e^-e^+ \to 4j$ process}
\label{sec:fullhad}
The event simulation follows similar to that of the semi-leptonic case discussed in the previous section. To quantify the size of the interference between the resonant $W^-W^+$ contribution and the Non-$W^-W^+$ (QCD and electroweak) topologies, we first generate dedicated event samples using a diagram filter in MG5. The filter removes the resonant $W^-W^+$ amplitudes, thereby yielding a sample that includes only the non-resonant QCD and electroweak ($VV$, with $V \in \{Z, \gamma\}$) contributions. This Non-$W^-W^+$ sample effectively serves as the background to the fully hadronic $W^-W^+$ signal in $e^-e^+ \to 4j$ production. The inclusive $4j$ sample, on the other hand, contains all gauge-invariant diagrams contributing to the process, including $WW$, $ZZ$, QCD, and their mutual interference, as schematically illustrated in Fig.~\ref{fig:fm_4jet}.

At $\sqrt{s} = 250~\text{GeV}$, the leading-order cross sections obtained are 
$\sigma_{WW} = 3.46~\text{pb}$, 
$\sigma_{4j} = 4.19~\text{pb}$, and 
$\sigma_{\mathrm{Non}\text{-}WW} = 0.63~\text{pb}$. 
The interference between the $W^-W^+$ and Non-$W^-W^+$ amplitudes is found to be small, approximately $0.1~\text{pb}$, corresponding to about $2.3\%$ of the total $e^+e^- \to 4j$ cross section when contributions up to $\mathcal{O}(\alpha_{\mathrm{EW}}^4 + \alpha_{\mathrm{EW}}^3\alpha_S + \alpha_{\mathrm{EW}}^2\alpha_S^2)$ are included.

To further characterize the dominant background topologies and to study the performance of event classification techniques, we simulate four distinct samples: (i) the inclusive $4j$ process containing all gauge-invariant contributions, (ii) the pure electroweak $W^-W^+$ process, representing the signal, (iii) the pure electroweak $ZZ \to 4j$ process and  (iv) the QCD-like $4j$ process. 

\begin{figure*}[!thb]
	\centering
	\includegraphics[width=0.49\textwidth]{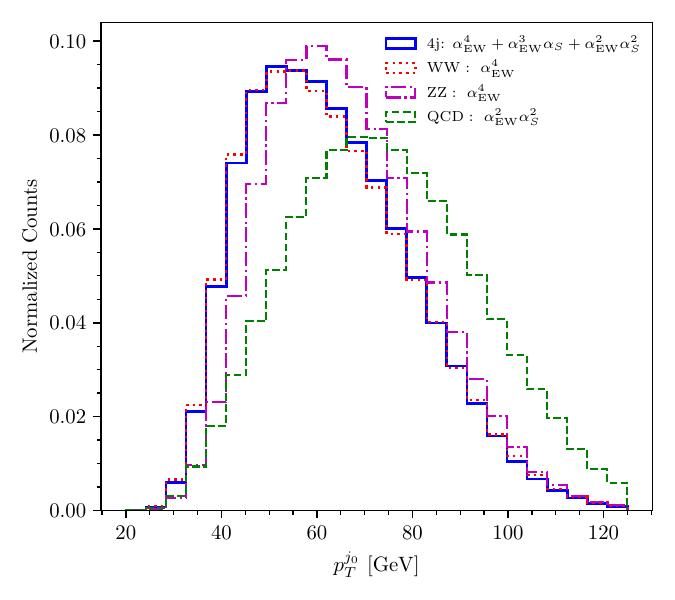}
	\includegraphics[width=0.49\textwidth]{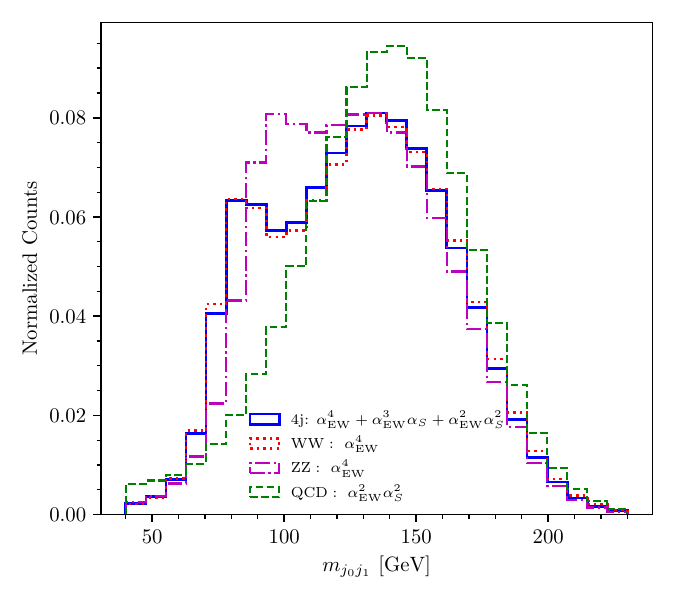}
	\caption{Detector level distribution of transverse momenta of hardest jet and invariant mass of pair of hardest and sub-leading jet for four different processes at $\sqrt{s}=250$ GeV. The distribution are obtained for $e^-e^+ \to 4j$ process along with di-boson ($W^-W^+, ZZ$) and QCD production. }
	\label{fig:njetmj1j3}
\end{figure*}

Figure~\ref{fig:njetmj1j3} shows the transverse momenta of hardest jet and the invariant mass distributions of leading jet pairs for the different process categories. The shown kinematic observables present a similar distribution among signal and two major backgrounds. We use these observables along with various other kinematic observables  to distinguish the $W^-W^+$ signal from the dominant $ZZ$ and QCD backgrounds using BDT, as discussed below.

\subsection{Event selection with boosted decision trees}
To distinguish between $WW$, $ZZ$, and QCD-induced four-jet topologies, we employ a gradient-boosted decision tree multi-class classifier. The input feature set, detailed in Table~\ref{tab:evttag}, comprises kinematic observables constructed from the four leading jets: transverse momenta ($p_T^{j_i}$, $i \in \{0,1,2,3\}$), pseudorapidities ($\eta_{j_i}$), azimuth orientation ($\phi_{j_i}$), dijet invariant masses ($m_{j_a j_b}$), dijet transverse momentum sums ($p_T^{j_a j_b}$), angular separations ($\Delta R_{j_a j_b}$), the scalar sum of all jet transverse momenta ($H_T$), the four-jet invariant mass ($m_{4j}$), and jet charge observables ($Q_J^\kappa$) computed with varying $\kappa$ exponents.

\begin{table*}[!htb]
	\centering
	\renewcommand{\arraystretch}{1.5}
	\caption{\label{tab:evttag} List of observables used as an input features for boosted decision tree in order to classify signal vs. background events. The features are derived at $\sqrt{s}=250$ GeV at the detector (\textsc{Delphes}) level.}
	\begin{tabular*}{1.0\textwidth}{@{\extracolsep{\fill}}cccc@{}}
		\hline
		Features& Description & Features & Description  \\
		\hline
		$p_T^{j_i},~i \in \{0,\dots,3\}$& $p_T$ of four leading jets & $\eta_{j_i}$ & Pseudorapidity of jets
		\\
		$\phi_{j_i}$ & Azimuth orientation of jets & $p_T^j$ & Scalar sum of $p_T$ of all jets\\
		$m_j$&Invariant mass of all jets & $p_T^{j_aj_b}$ & Scalar sum of $p_T$ two jets  \\
		$m_{j_aj_b}$ & Invariant mass of two jets & $\Delta R_{j_aj_b}$ & Geometric distance\\
		$Q_J^k$ & Jet charge with different $k$ &  & \\
		\hline
	\end{tabular*}
\end{table*}

\begin{figure*}[!htb]
	\centering
	\includegraphics[width=0.49\textwidth]{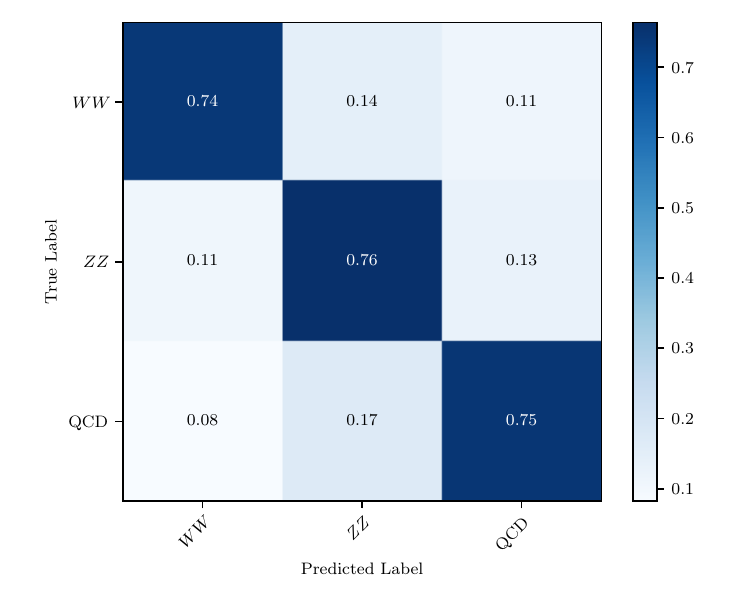}
    \includegraphics[width=0.49\textwidth]{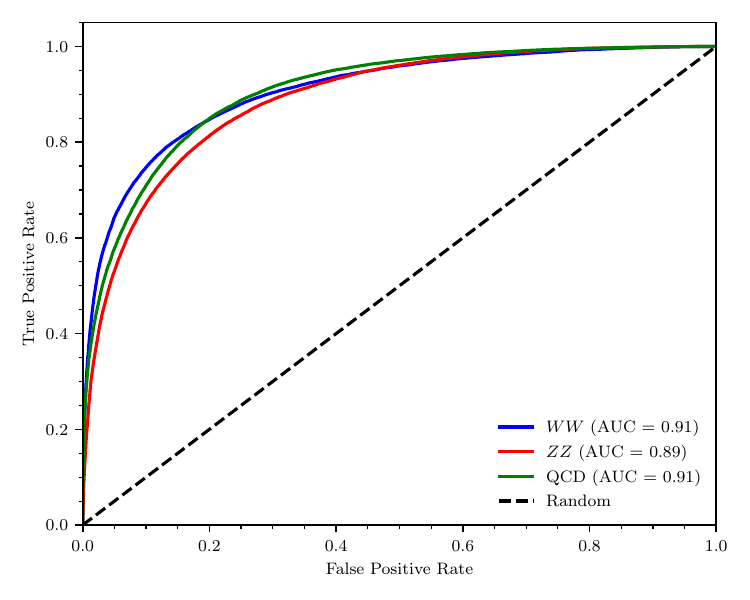}
	\caption{Confusion matrix and ROC curves to highlight the efficiency of network to classify signal events ($W^+W^-$) from the $ZZ$ and QCD backgrounds.}
	\label{fig:roc}
\end{figure*}

Hyperparameter tuning is performed via exhaustive grid search with three-fold cross-validation, minimizing the multi-class logarithmic loss. The optimal configuration consists of: maximum tree depth $= 8$, learning rate $\eta = 0.05$, number of estimators $= 500$, and sub-sample ratio $= 0.7$. Input features are preprocessed using mean imputation for missing values and standard scaling to zero mean and unit variance. The model employs the softmax objective function to output calibrated class probabilities for the three event categories. 

The trained classifier achieves robust discrimination between the $WW$ signal and backgrounds ($ZZ$ and QCD), as quantified by the confusion matrix and one-versus-rest receiver operating characteristic (ROC) curves presented in Fig.~\ref{fig:roc}. The confusion matrix demonstrates that $WW$ events are correctly identified with $74\%$ efficiency (true positive rate), while background contamination remains modest: $12\%$ of $ZZ$ events are misclassified as $WW$ and $8\%$ dilution comes from QCD. The dominant confusion with $ZZ$ backgrounds is expected, as both processes yield diboson topologies with similar jet multiplicity and kinematic distributions, differing primarily in the invariant mass reconstructions of the decay products. It results in an improved signal purity $(P = N_S/(N_S+N_B))$ that must be accounted for in subsequent sensitivity analyses. In particular, the signal purity $P$ increases from $\approx 90\%$ to $99\%$ on using multi-variate classifier. 
\begin{figure*}[!thb]
	\centering
	\includegraphics[width=0.32\textwidth]{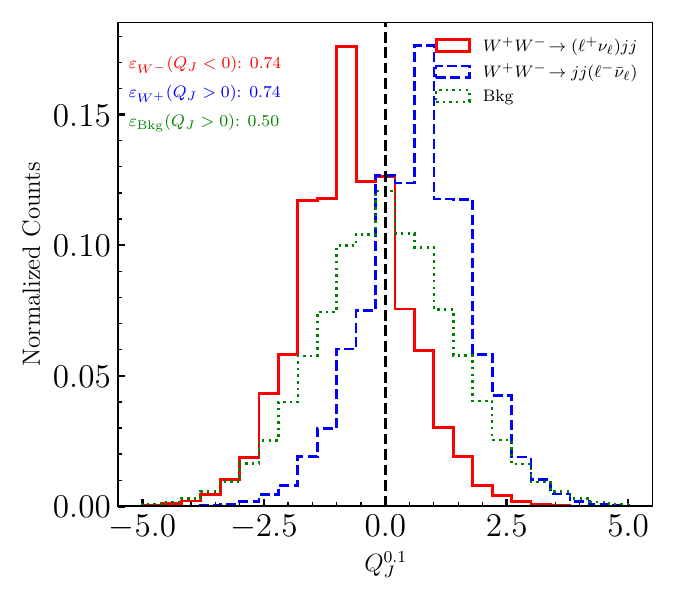}
	\includegraphics[width=0.32\textwidth]{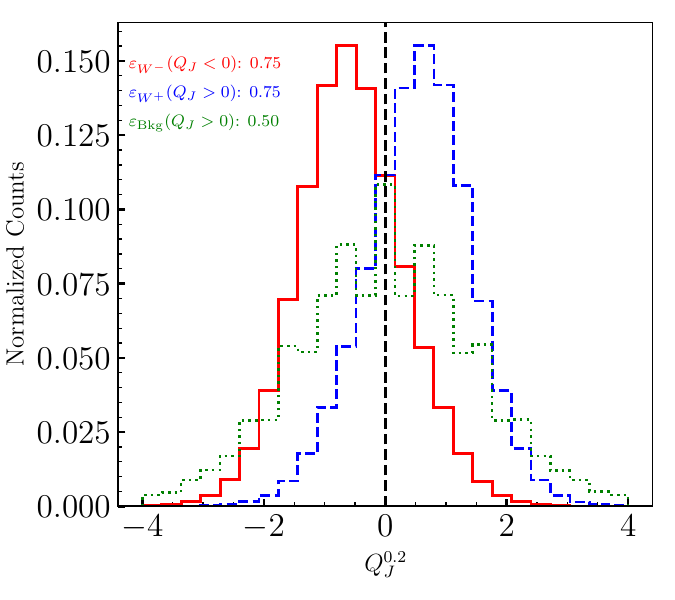}
	\includegraphics[width=0.32\textwidth]{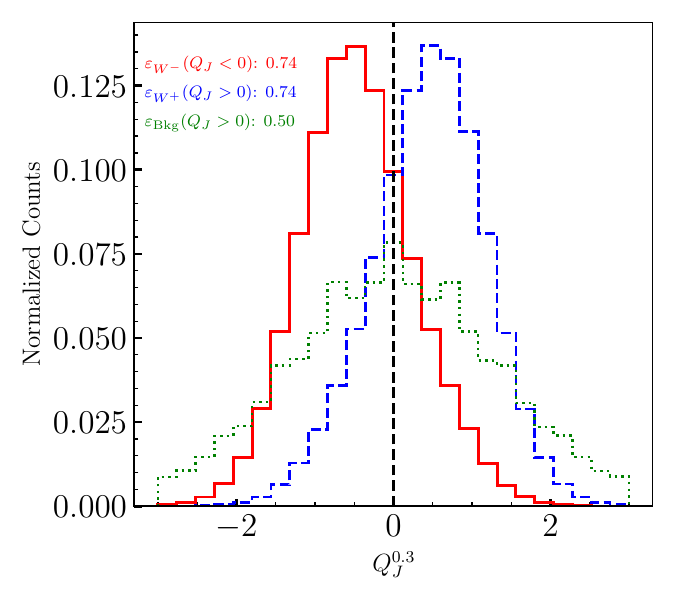}
	\includegraphics[width=0.32\textwidth]{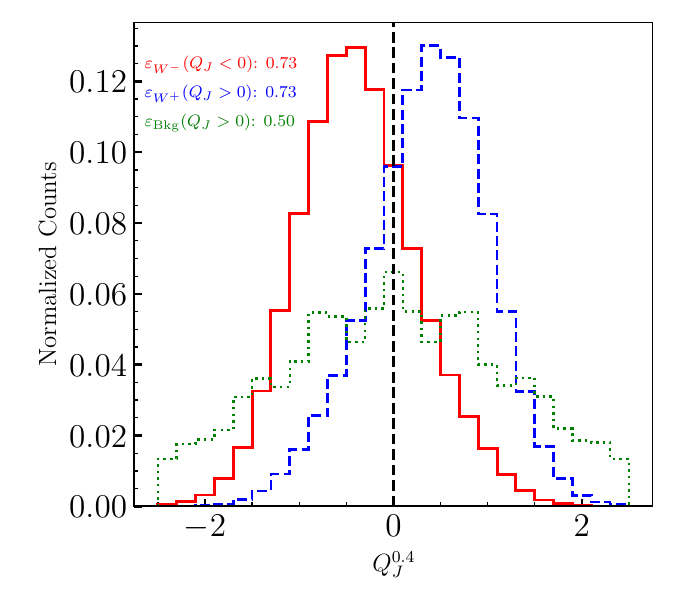}
	\includegraphics[width=0.32\textwidth]{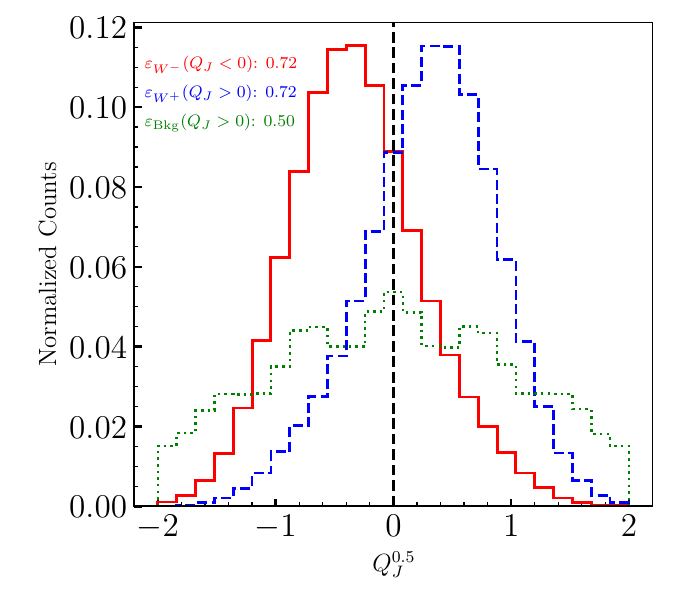}
	\includegraphics[width=0.32\textwidth]{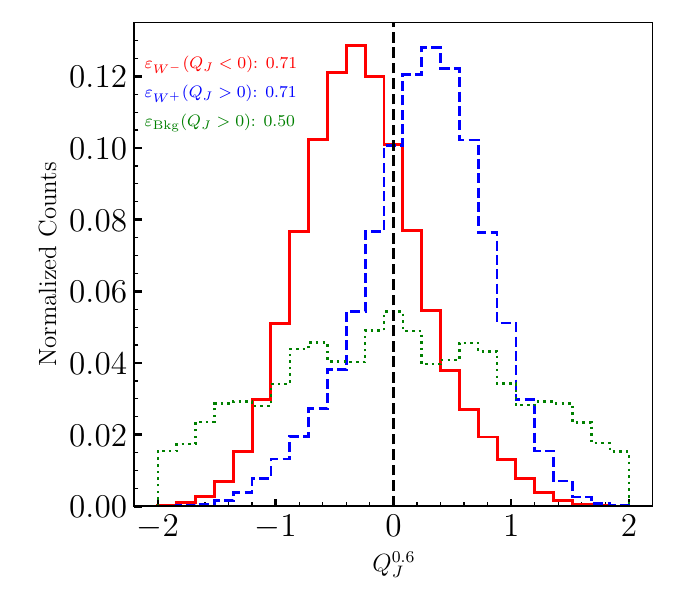}
	\includegraphics[width=0.32\textwidth]{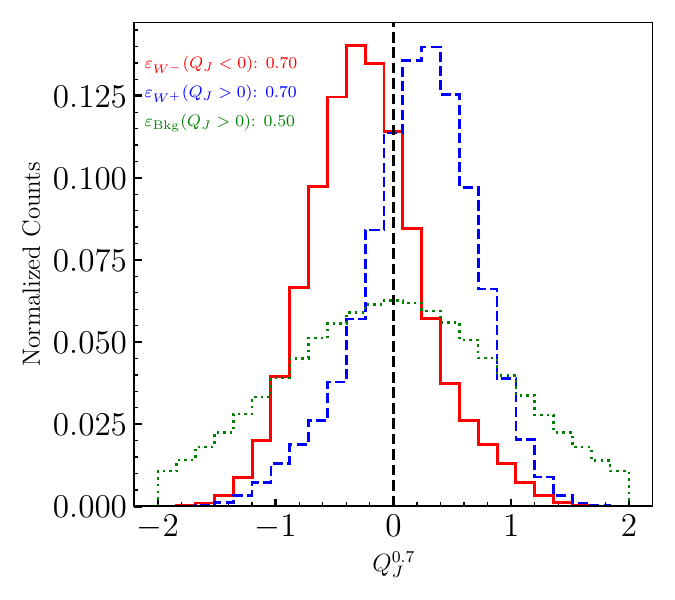}
	\includegraphics[width=0.32\textwidth]{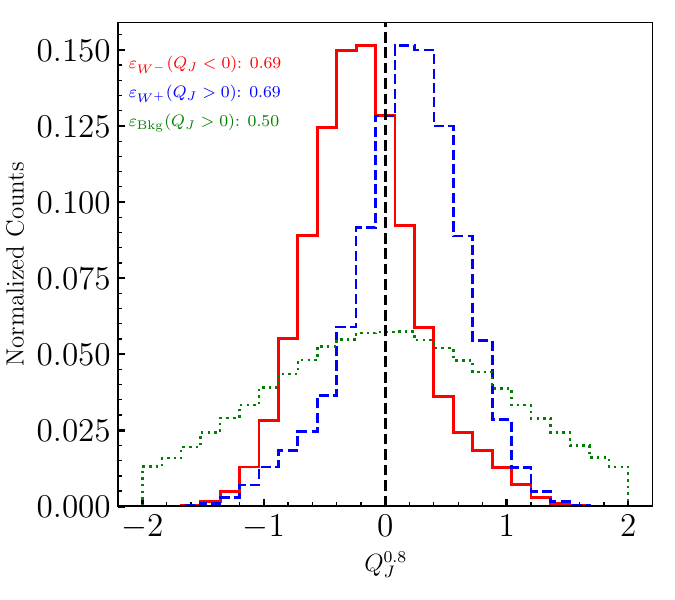}
	\includegraphics[width=0.32\textwidth]{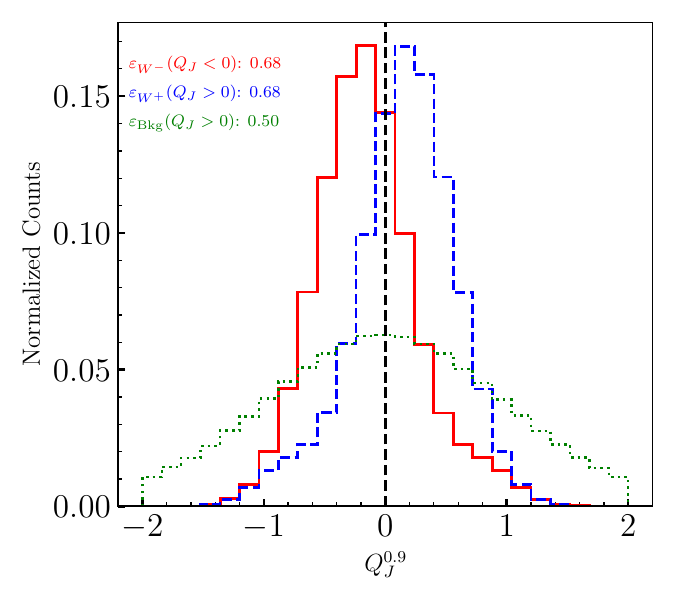}
	\caption{Normalized distribution of jet charge variable for two different semi-leptonic decay of $WW$ di-boson process and background events for four jet events. The distribution are obtained at the level of detector implemented in \texttt{Delphes} at $\sqrt{s}=250$ GeV.}
	\label{fig:jc}
\end{figure*}

The \emph{one-versus-rest} ROC analysis quantifies the $WW$ signal-background separation power through the area under the curve (AUC). The $WW$ classifier achieves $\mathrm{AUC}_{WW} = 0.91$, indicating excellent discrimination against the combined $ZZ$ and QCD backgrounds. This high AUC value reflects the classifier's ability to exploit the characteristic dijet mass structure of $W$ boson decays ($m_W \approx 80$~GeV) in conjunction with angular observables and jet charge asymmetries. For comparison, the background classifiers yield $\mathrm{AUC}_{ZZ} = 0.89$ and $\mathrm{AUC}_{\mathrm{QCD}} = 0.91$, demonstrating that the multi-class approach successfully disentangles all three topologies.

\subsection{Reconstruction of oppositely charged two $W$ bosons}
\label{sec:jetcharge}
Once the $WW$-like events are tagged, the next question to tackle is to figure out which pair of jets is more closely $W^-$ or $W^+$ aligned. This $W$ tagging is necessary in order to reconstruct the polarization and spin correlations of two $W$ bosons. From the kinematic consideration, it becomes extremely difficult to tag the jet pair to respective charge as the hadronization of decayed quarks from two oppositely charged $W$ bosons are similar leading to similar jet kinematic structure. 
\begin{figure*}[!htb]
	\centering
	\includegraphics[width=0.49\textwidth]{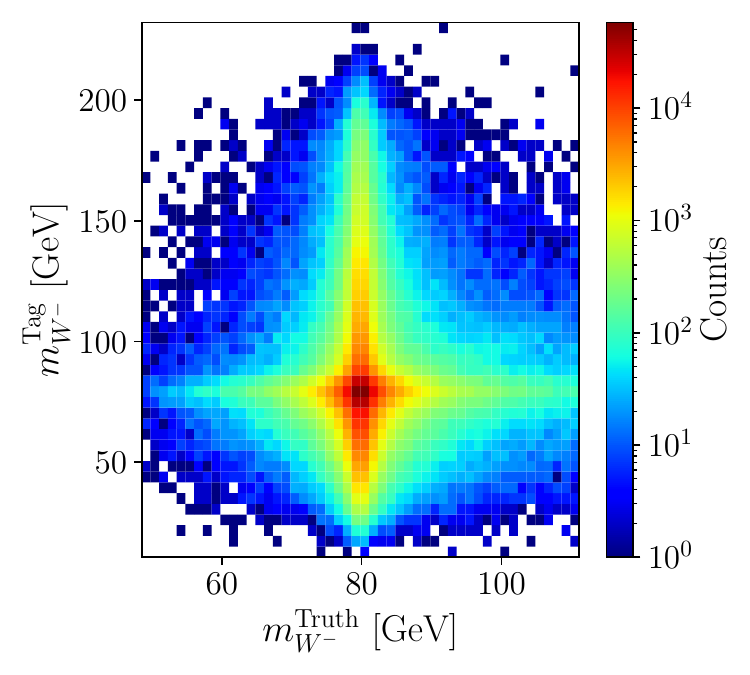}
	\includegraphics[width=0.49\textwidth]{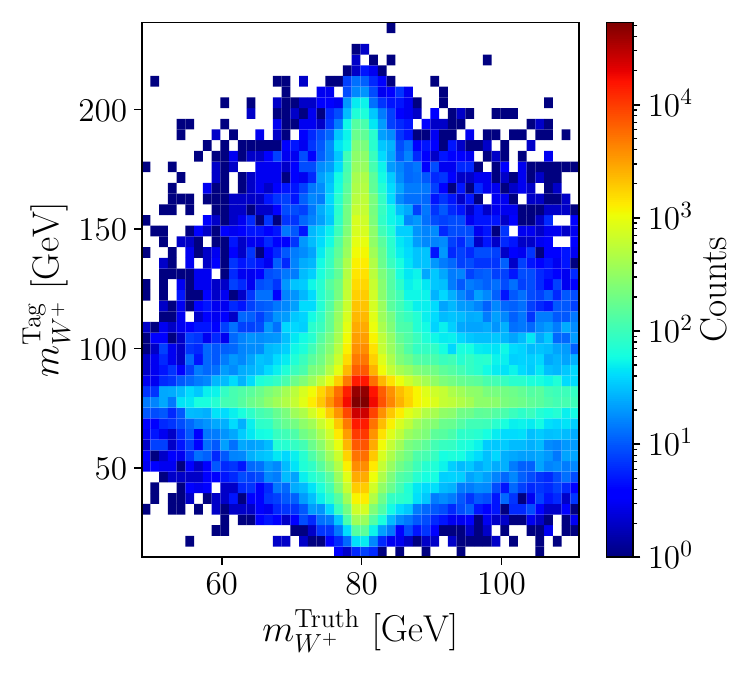}
	\caption{Two dimensional distribution of invariant mass of $W$ bosons at the level of reconstruction and parton level. The distribution are obtained for full-hadronic decay of $W^-W^+$ di-boson process at $e^-e^+$ collider at $\sqrt{s}=250$ GeV.}
	\label{fig:wmass}
\end{figure*}

To discriminate between jets originating from $W^-$ and $W^+$ decays, we employ the $p_T$-weighted jet charge observable ($Q_J$) defined in Eq.~\eqref{eq:jet_charge}. To quantify the discriminating power of $Q_J^\kappa$, we simulate two semi-leptonic $e^-e^+ \to W^-W^+ \to jj(l^\pm\nu_l)$ processes, along with the inclusive $e^-e^+ \to 4j$ process. The latter is generated with a diagram filter that removes the $W^+W^-$ resonant amplitudes, ensuring that only non-resonant QCD and electroweak ($VV$, $V\in\{Z,\gamma\}$) topologies contribute and act as background to full-hadronic channel. We focus on the semi-leptonic channels to avoid the combinatorial ambiguity associated with pairing jets in the fully hadronic final state. In Fig.~\ref{fig:jc}, we show the normalized distributions of $Q_J^\kappa$ for different values of $\kappa \in \{0.1,\cdots,0.9\}$ across the three processes. A clear charge asymmetry is observed between the two semi-leptonic $WW$ channels, while the $4j$ background remains symmetric around zero. This symmetry indicates that $Q_J$ is an effective discriminator only for resonant $WW$ topologies. The best tagging performance is achieved for $\kappa=0.2$, yielding a tagging efficiency of approximately $0.75$. Further, in the fully hadronic $e^-e^+\to 4j$ case, the other challenge arises from the combinatorial ambiguity in pairing jets into $W$ candidates. Let $J_i \in \{J_1,J_2,J_3,J_4\}$ denote the four leading jets. Three distinct pairings can be formed: $P_1=\{(J_1,J_2),(J_3,J_4)\}$, $P_2=\{(J_1,J_3),(J_2,J_4)\}$, and $P_3=\{(J_1,J_4),(J_2,J_3)\}$. We select the pairing for which the product of the two jet charges is negative. If multiple pairings satisfy this condition, the one minimizing $||m_{Ji}-m_W| + |m_{Jj}-m_W||$ is chosen.

Thus, in the current work we employ $Q_J^{0.2}$, assuming the same optimal $\kappa$ value applies as in the semi-leptonic case, since the observable primarily probes jet substructure, which is expected to be similar in both topologies. The jet pair with negative (positive) total charge is identified as the $W^-$ ($W^+$) candidate.

Figure~\ref{fig:wmass} shows the correlation between the reconstructed and parton-level invariant masses of the charged $W$ bosons for the fully hadronic $e^-e^+\to 4j$ process. For roughly $80\%$ of the events, the reconstructed $W$ mass deviates from the parton-level value by less than $20~\mathrm{GeV}$. A vertical condensation pattern is visible, which reflects fluctuations in the reconstructed jet energies. A similar feature is also observed in the semi-leptonic channel, suggesting that the spread originates mainly from hadronization and detector effects rather than from incorrect jet pairing. Once the charged $W^\pm$ bosons are successfully tagged, we proceed to identify the flavor of their daughter jets following a similar tagging procedure described for semi-leptonic case in the previous section.

\section{The $e^-e^+ \to \ell^-\ell^+\slashed{E}$ process}
\label{sec:full-lep}
\begin{figure*}[!htb]
	\centering
	\includegraphics[width=0.49\textwidth]{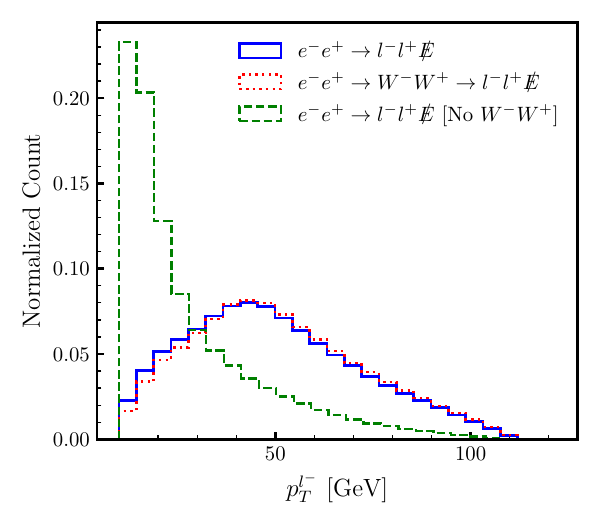}
	\includegraphics[width=0.49\textwidth]{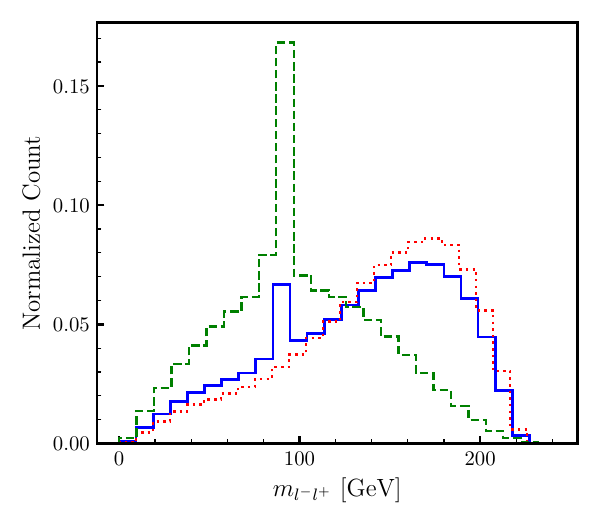}
	\caption{Detector level kinematic distribution of leptons and missing neutrino for $\ell^-\ell^+\slashed{E}$ events at $\sqrt{s}=250$ GeV $e^-e^+$ collider. The distribution are obtained at leading order for three different processes viz. i) complete events $e^-e^+ \to \ell^-\ell^+ \slashed{E}$, ii) signal events $e^-e^+ \to W^-W^+ \to \ell^-\ell^+ \slashed{E}$, and iii) background process devoid of $W^-W^+$ amplitudes.}
	\label{fig:dilepkin}
\end{figure*}
At leading order (LO), the production of final states comprising two oppositely charged leptons and missing transverse energy at an electron-positron collider arises from several distinct classes of Feynman diagrams, as illustrated in Fig.~\ref{fig:fm_dilep}. The primary signal process corresponds to $W^-W^+$ production, where the intermediate states involve both $\gamma/Z$-mediated $s$-channel diagrams and $\nu$-mediated $t$-channel exchange. These diagrams contribute dominantly to the electroweak di-boson topology and serve as the principal focus of our analysis.

In contrast, the dominant irreducible backgrounds arise from vector boson fusion (VBF) processes, which involve $t$-channel exchange of gauge bosons resulting in final-state leptons and neutrinos, as depicted in the bottom row of Fig.~\ref{fig:fm_dilep}. In addition, there exists a continuum background arising from non-resonant contributions that mimic the same final state. To enhance the sensitivity to the di-boson signal and reduce contamination from background topologies, we simulate three distinct event samples within our Monte Carlo simulation pipeline. The first sample includes the complete set of LO amplitudes contributing to the $e^-e^+ \to \ell^-\ell^+\slashed{E}$ final state. The second sample isolates only the pure signal processes corresponding to $W^-W^+$ production. The third sample comprises background-only contributions, excluding all diagrams associated with the di-boson topology. These samples are generated using the diagram filtering capabilities of the MG5 framework.

It is worth emphasizing that the dimension-six operators introduced in the previous section not only modify the $W^-W^+$ production amplitudes  but also contribute anomalously to the VBF processes. As such, VBF topologies may serve as an independent and complementary probe of new physics effects. Nevertheless, in the present study, we restrict our attention to the di-boson rich phase space, while a detailed investigation of VBF-sensitive observables is deferred to future work.

In Fig.~\ref{fig:dilepkin}, we present the normalized detector-level distributions of transverse momenta of negatively charged lepton ($p_T^{\ell^-}$) and invariant mass of di-lepton ($m_{\ell^-\ell^+}$). Both the observables exhibit significant discriminating power between the signal and background processes. Specifically, the $p_T^{\ell^-}$ distribution for background events exhibits a peak near $10$ GeV, whereas the corresponding distribution for signal events peaks at approximately $50$ GeV. Similarly, in the $m_{\ell^-\ell^+}$ distribution, the background contribution features a pronounced peak around $90$ GeV consistent with $Z$-boson mediated process, while the signal exhibits a broader distribution with a peak around $160$ GeV, reflecting the kinematic features of $W^-W^+$ production.

We use cut-based analysis to select the signal rich phase space. Events with two oppositely charged leptons and missing transverse momenta, $\slashed{p}_T \ge 10$ GeV are selected with cuts
$$p_T^{\ell^-} \ge 25~\mathrm{GeV},~~m_{\ell^-l^+} \ge 130.0~\mathrm{GeV}.$$
Once the events are selected, the next hurdle is the reconstruction of 4-momenta of two missing neutrinos. It is required in order to compute the spin polarizations and correlations asymmetries at the rest frame of mother $W$ bosons.

\subsection{Reconstruction of two missing neutrinos}
\label{sec:recomet}
\begin{figure*}[!htb]
	\centering
	\includegraphics[width=0.99\textwidth]{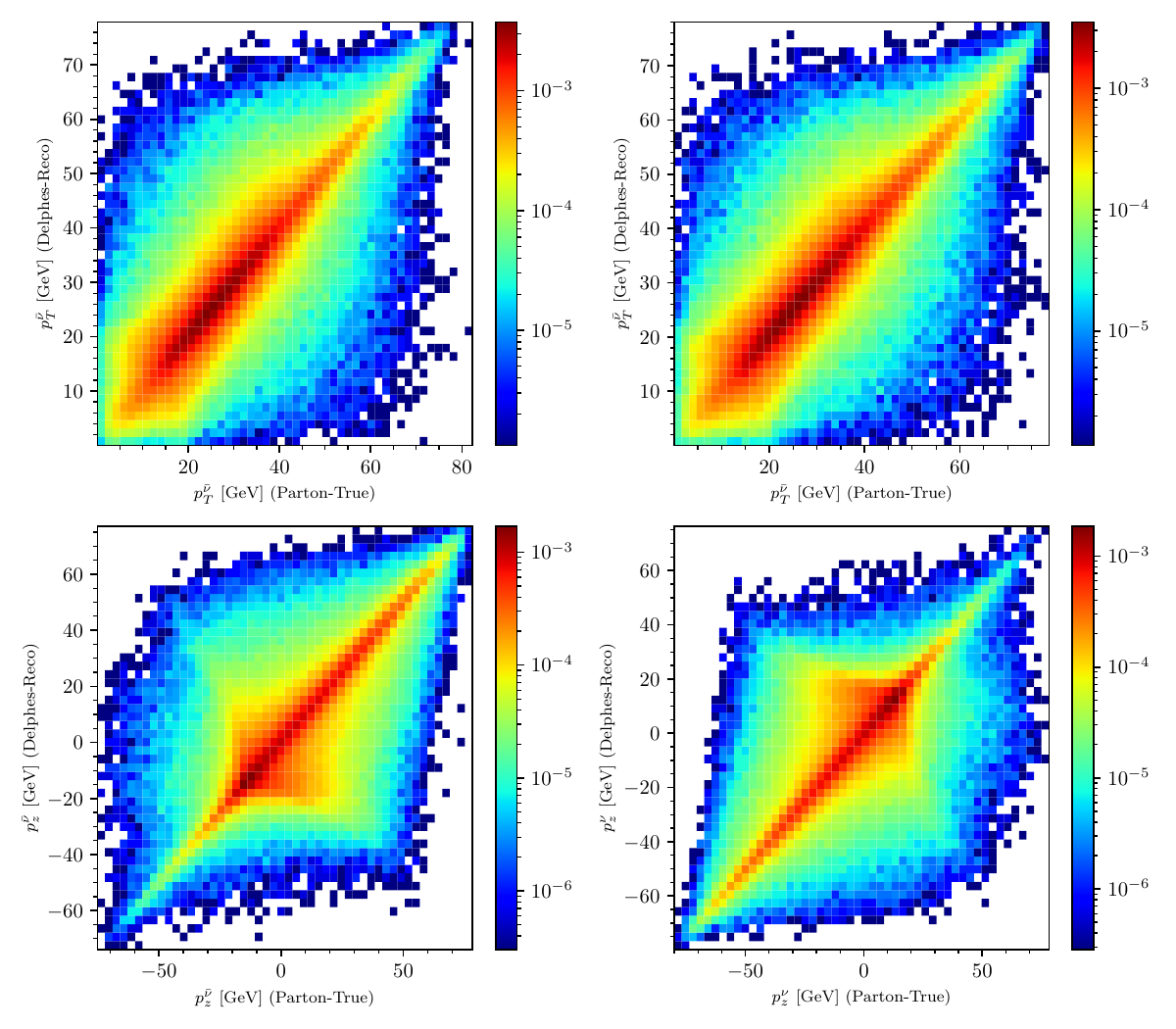}
	\caption{\label{fig:twoneutrino}
		Two-dimensional correlation between the true parton-level and reconstructed
		\textsc{Delphes}-level neutrino momenta. The top row corresponds to
		transverse ($p_T$) and the bottom row to longitudinal ($p_z$) momenta. 
		The distributions are obtained for
		$e^-e^+ \to \ell^-\ell^+\slashed{E}$ final state at $\sqrt{s}=250$ GeV with
		initial beam polarization of $P_{e^-}=-80\%$ and $P_{e^+}=+30\%$.}
\end{figure*}
The presence of two undetected neutrinos in the final state of $W^-W^+$ events presents a significant challenge for fully reconstructing the event kinematics. We consider the process
\begin{align}
	e^-(p) + e^+(\bar{p}) \to &W^-(p_1) + W^+(p_2), \nonumber\\
	&W^- \to \ell^-(l_1) + \bar{\nu}_\ell(k_1), \\
	&W^+ \to \ell^+(l_2) + \nu_\ell(k_2), \nonumber
\end{align}
where $l_1$ and $l_2$ are the four-momenta of the charged leptons, and $k_1$ and $k_2$ are the four-momenta of the antineutrino and neutrino, respectively. The reconstruction of two missing neutrino momenta is performed using the analytical method described in Ref.~\cite{Hagiwara:1986vm}, which we briefly summarize here for completeness.

In the center-of-mass frame, the energy of each $W$ boson is fixed to the beam energy $E_b = \sqrt{s}/2$. This constraint, combined with the massless neutrino assumption, yields $|\mathbf{k}_1|^2 = (E_b - l_1^0)^2 \equiv L$, where $l_1^0$ is the energy of the charged lepton from the $W^-$ decay. The on-shell condition for the $W$ boson provides two additional constraints:
\begin{align}
	m_{W^-}^2 &= (l_1 + k_1)^2, \\
	m_{W^+}^2 &= (l_2 + k_2)^2,
\end{align}
which can be rewritten as
\begin{align}
	\label{eq:M}
	\mathbf{l}_1\cdot\mathbf{k}_1 &= E_b l_1^0 - {l_1^0}^2 - \tfrac{1}{2}m_W^2 + \tfrac{1}{2}m_\ell^2 \equiv M, \\
	\label{eq:N}
	\mathbf{l}_2\cdot\mathbf{k}_1 &= -E_b l_2^0 - \mathbf{l}_1\cdot\mathbf{l}_2 + \tfrac{1}{2}m_W^2 + \tfrac{1}{2}m_\ell^2 \equiv N.
\end{align}
For non-collinear charged lepton momenta, the anti-neutrino three-momentum can be expressed as a linear combination of the charged lepton momenta and their cross product:
\begin{equation}
	\mathbf{k}_1 = a\,\mathbf{l}_1 + b\,\mathbf{l}_2 + c\,(\mathbf{l}_1 \times \mathbf{l}_2).
\end{equation}
The linear constraints in Eq.~\eqref{eq:M} and \eqref{eq:N} determine the coefficients $a$ and $b$:
\begin{align}
	a|\mathbf{l}_1|^2 + b(\mathbf{l}_1\cdot\mathbf{l}_2) &= M, \\
	a(\mathbf{l}_1\cdot\mathbf{l}_2) + b|\mathbf{l}_2|^2 &= N.
\end{align}
The remaining coefficient $c$ is then obtained from the energy constraint:
\begin{equation}
	c^2 = \frac{L - a^2|\mathbf{l}_1|^2 - b^2|\mathbf{l}_2|^2 - 2ab(\mathbf{l}_1\cdot\mathbf{l}_2)}{|\mathbf{l}_1 \times \mathbf{l}_2|^2}.
\end{equation}
This yields a twofold ambiguity in the neutrino solution corresponding to the two signs of $c$. To resolve this ambiguity and account for the finite width of the $W$ boson, we adopt the following procedure:
\begin{enumerate}
	\item Generate a set of 1,000 trial $W$ boson masses sampled from a Gaussian distribution centered at the nominal $W$ mass $m_W = 80.4$~GeV with standard deviation $\sigma = 2.1$~GeV, in addition to the nominal value. The $\sigma = 2.1$~GeV is considered by noting the Breit-Wigner width of $W$ boson.
	\item For each trial mass, compute both solutions corresponding to $c = \pm\sqrt{c^2}$.
	\item Among all valid solutions, select the one that minimizes the reconstructed $W$ mass difference:
	\begin{equation}
		\Delta m_W = \left| \sqrt{(l_1^0 + k_1^0)^2 - |\mathbf{l}_1 + \mathbf{k}_1|^2} - m_W^{\text{true}} \right|.
	\end{equation}
\end{enumerate}
The second neutrino momentum is then obtained from momentum conservation:
\begin{equation}
	\mathbf{k}_2 = -(\mathbf{k}_1 + \mathbf{l}_1 + \mathbf{l}_2).
\end{equation}
The reconstruction algorithm is applied to both parton-level events (using generator-level four-momenta) and detector-level events after simulation with \textsc{Delphes}. Events for which no physical solution exists are discarded.

The goodness of the neutrino reconstruction is illustrated in Fig.~\ref{fig:twoneutrino}, where the true parton-level momenta are plotted against the reconstructed \textsc{Delphes}-level momenta. A clear diagonal band is visible, indicating a strong correlation. Quantifying this with the Pearson coefficient, we find for the transverse momentum: $r(p_T)=0.80$ for $\bar{\nu}$ and $r(p_T)=0.82$ for $\nu$. For the longitudinal component, the correlation is lower: $r(p_z)=0.71$ for $\bar{\nu}$ and $r(p_z)=0.70$ for $\nu$. This reflects
the fact that the longitudinal momentum is less constrained at colliders, leading to a broader reconstruction. Overall, the \textsc{Delphes}-level reconstruction captures the transverse momentum well, while the longitudinal component is smeared more strongly.

\begin{figure*}[!t]
	\centering
	\includegraphics[width=0.32\textwidth]{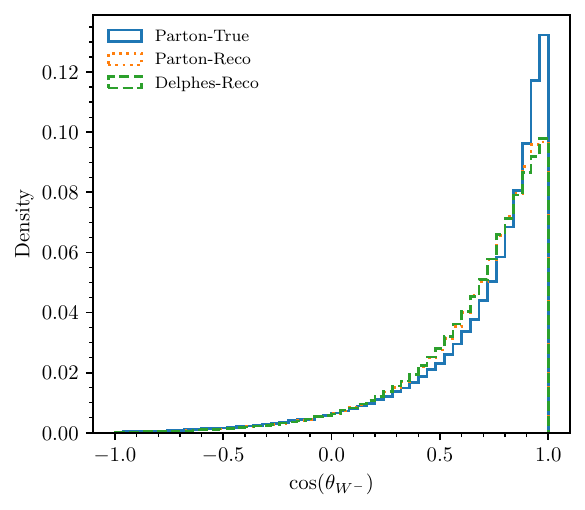}
	\includegraphics[width=0.32\textwidth]{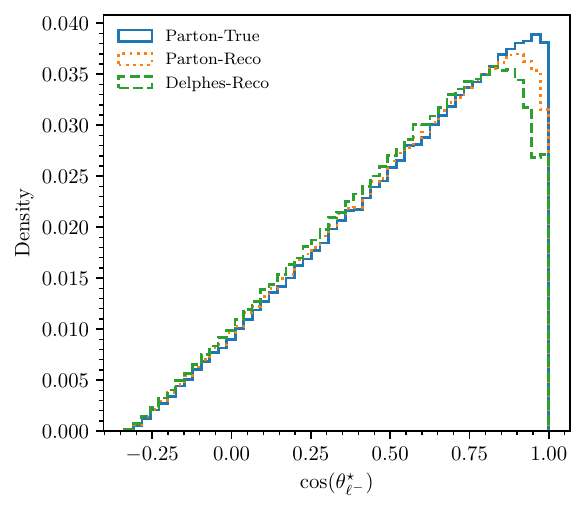}
	\includegraphics[width=0.32\textwidth]{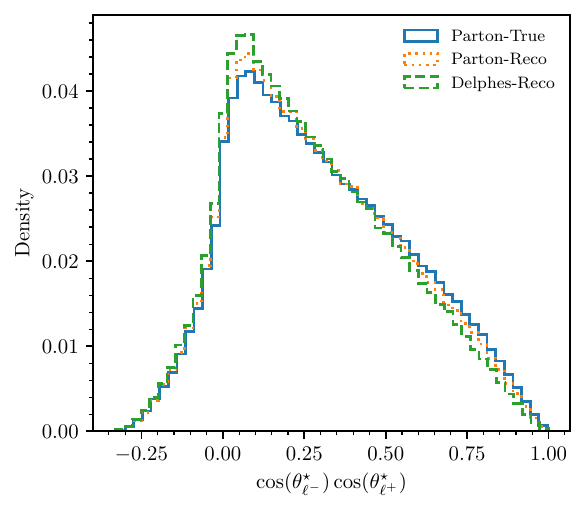}
	\caption{\label{fig:angles_compare_lep}Normalized distribution of angular distribution of $W^-$ boson along with final leptons at the level of true parton (solid blue)  and reconstruction at the parton (dotted red) and \textsc{Delphes} (dashed green) level. The distribution are obtained for $e^-e^+ \to \ell^-\ell^+\slashed{E} $ final state at $\sqrt{s}=250$ GeV with initial beam polarization of $P_{e^-}=-80\%$ and $P_{e^+}=+30\%$.  }
\end{figure*}

Further, the goodness of the reconstruction is demonstrated by comparing the angular distributions of the reconstructed $W^-$ boson and the final-state charged leptons in Fig.~\ref{fig:angles_compare_lep}. We show
$\cos\theta_{W^-}$ in the lab frame, and $\cos\theta_{\ell^-}^\star$, and $\cos\theta_{\ell^-}^\star\cos\theta_{\ell^+}^\star$ in the rest frame of the $W$ boson. These observables are directly sensitive to the $W$ polarization and to the spin correlations between the two $W$ bosons. The true parton-level, parton-level reconstructed, and \textsc{Delphes}-level reconstructed distributions agree well in shape and normalization, showing
that the reconstruction preserves the angular kinematics relevant for polarization and spin-correlation measurements.
	
\section{SMEFT analysis}
\label{sec:results}

This section discusses the methodology employed to constrain the anomalous couplings. As discussed in the above section, $16$ polarization and $64$ spin-correlation asymmetries exist in the case of pair produced $W$
boson.  In order to effectively constrain the anomalous couplings, we have categorized all observables into four distinct intervals of $\cos\theta_{W^-}$ as
\begin{align}
    C_n \equiv -1.0 + \frac{n-1}{2} \le \cos\theta_{W^-} < -1 + \frac{n}{2},
\end{align}
where $\theta_{W^-}$ represents the production angle of the reconstructed $W^-$ boson in the lab frame. Within each bin, we have identified $1$ cross-section along with 80 asymmetries resulting in a total of $648$ observables. These observables have been computed for the SM and several benchmark anomalous points. These values are used for numerical fitting to obtain a semi-analytical expression of observables as a function of anomalous couplings. For cross~section, which is a CP-even observable, the following parametric function is used,
\begin{align}
    \sigma(\{c_i\}) &= \sigma_0 + \sum_{i=1}^3\sigma_ic_i + \sum_{j=1}^5\sigma_{jj}c_j^2
    + \sum_{i> j}^3\sigma_{ij}c_ic_j + \sigma_{45}c_4c_5,
    \label{eqn:cpeven}
\end{align}
where couplings $c_i\in \{c_1\equiv c_{WWW},c_2\equiv c_W,c_3\equiv c_B\}$ corresponds to CP-even and
$c_i\in\{c_4\equiv c_{\widetilde{W}WW},c_5\equiv c_{\widetilde{W}}\}$ corresponds to CP-odd WCs. In the case of asymmetries, the denominator is cross~section, while the numerator corresponds to difference in cross~section. While fitting the asymmetries $(\mathcal{A})$, we construct a cross~section weighted function, with numerator is defined as
$\Delta\sigma = \mathcal{A} \times \sigma$ and the denominator is simply the cross~section. The numerator of the CP-odd asymmetries are fitted using the
function,
\begin{equation}
    \Delta\sigma(\{c_i\}) = \sum_{i=4}^5\sigma_ic_i + \sum_{i=1}\sigma_{i4}c_ic_4+ \sum_{i=1}\sigma_{i5}c_ic_5,
    \label{eqn:cpodd}
\end{equation}
while the CP-even numerator are parameterize using Eq.~(\ref{eqn:cpeven}).

Further, each $\cos\theta_{W^-}$ bin contains $80$ asymmetries extracted from the same event sample and sharing the same reconstructed final-state kinematics. They can therefore be statistically correlated: a fluctuation in the event
selection or in the $W$ reconstruction propagates into several polarization and spin-correlation asymmetries simultaneously. A diagonal-error $\chi^2$ would therefore mis-estimate the constraining power of the fit. For a vector of $80$ asymmetries $\mathbf{A} = (\mathcal{A}_1,\dots,\mathcal{A}_{80})$, a set of WCs $c$, and a beam-polarization configuration $(\mp P_{e^-},\pm P_{e^+})$, we adopt the full covariance-based quadratic form
\begin{equation}
  \Delta\chi^2\!\left(\mathbf{A},c\right) = \bigl(\mathbf{A}(c) - \mathbf{A}(0)\bigr)^{\!T}\,
    \mathrm{M}^{-1}\, \bigl(\mathbf{A}(c) - \mathbf{A}(0)\bigr),
  \label{eqn:chisq_cov}
\end{equation}
where $\mathbf{A}(0)$ is the SM prediction, $\mathbf{A}(c)$ the SMEFT prediction, and $\mathrm{M}$ the $80\times 80$ covariance matrix of the SM asymmetries. When both beam polarizations are combined, the total $\Delta\chi^2$ is the sum of the two contributions, with $\mathrm{M}$ constructed separately for each polarization configuration. 

The covariance is estimated by drawing $N=1000$ independent random $40\%$ subsets from the SM event sample and computing the full asymmetry vector $\mathbf{A}^{(k)}$ for each subset $k$. The sample mean and covariance are
\begin{align}
  \bar{\mathcal{A}}_i &= \frac{1}{N}\sum_{k=1}^N \mathcal{A}_i^{(k)},
  \nonumber\\
  \mathrm{M}_{ij} &= \frac{1}{N-1}\sum_{k=1}^N
    \bigl(\mathcal{A}_i^{(k)} - \bar{\mathcal{A}}_i\bigr)
    \bigl(\mathcal{A}_j^{(k)} - \bar{\mathcal{A}}_j\bigr).
  \label{eqn:cov}
\end{align}
Since  each subset is drawn without replacement from a common pool, the sample variance is suppressed by the factor $(1-f)$ with $f=0.4$. We therefore rescale $\mathrm{M}\to\mathrm{M}/(1-f)$ to recover the true variance at the reference luminosity $\mathcal{L}_0$ (estimated from the simulation) of the pool. The statistical part of the covariance scales as $1/\mathcal{L}$, so we propagate $\mathrm{M}_{\rm stat}(\mathcal{L}) = \mathrm{M}_{\rm stat}\, (\mathcal{L}_0/\mathcal{L})$ to any target luminosity. A systematic contribution $\mathrm{M}_{\rm syst}$ is added separately and in the current work we consider a flat systematics blind to the correlations.

We have verified that the diagonal errors extracted from Eq.~(\ref{eqn:cov}) agree with the analytic expressions
\begin{equation}
  \delta\mathcal{A} = \sqrt{\frac{1-\mathcal{A}^2}{\mathcal{L}\sigma}
  + \epsilon_A^2},
\end{equation}
to within the statistical precision of the $1000$ subsets, where $\epsilon_A$ are the fractional systematic errors on $\mathcal{A}$. The full form of Eq.~(\ref{eqn:chisq_cov}) reduces to the diagonal approximation when $\mathrm{M}_{ij}\to\delta_{ij}\,\delta\mathcal{O}_i^2$. We also consider a contribution from cross~section to the overall sensitivity where the error in the estimation of cross~section is given by
\begin{align}
    \delta\sigma = \sqrt{\frac{\sigma}{\mathcal{L}} + (\epsilon_\sigma\sigma)^2},
\end{align}
with $\epsilon_\sigma$ being the systematic uncertainty in estimation of cross~section.

We begin by quantifying the impact of using initial beam polarization (IBP) which will be one of the technological upgrade in the upcoming $e^-e^+$ collider like ILC which planned to use polarized beams. In presence of longitudinal IBP, the polarized cross~section for $\gamma/Z$ mediated s-channel can be written as
\begin{align}
    \sigma(P_{e^-},P_{e^+}) = (1-P_{e^-}P_{e^+})\sigma_0\Big(1- P_{\mathrm{eff}}\,A_{LR}\Big),
\end{align}
where $\sigma_0$ is the unpolarized cross~section, $P_{\mathrm{eff}}$ is the effective polarization defined as
\begin{align}
    P_{\mathrm{eff}} = \frac{P_{e^-}-P_{e^+}}{1 - P_{e^-}P_{e^+}}
\end{align}
and the asymmetry in the polarized cross~section is $A_{LR} = (\sigma_{LR} - \sigma_{RL})/(\sigma_{LR}+\sigma_{RL}).$ For the $W$ or neutrino mediated $t$-channel, $\sigma_{RL} = 0$. In this work, we choose baseline potential of ILC with $(P_{e^-},\,P_{e^+}) = (\mp0.8,\,\pm0.3)$.

\begin{figure}[!htb]
    \centering
    \includegraphics[width=0.325\textwidth]{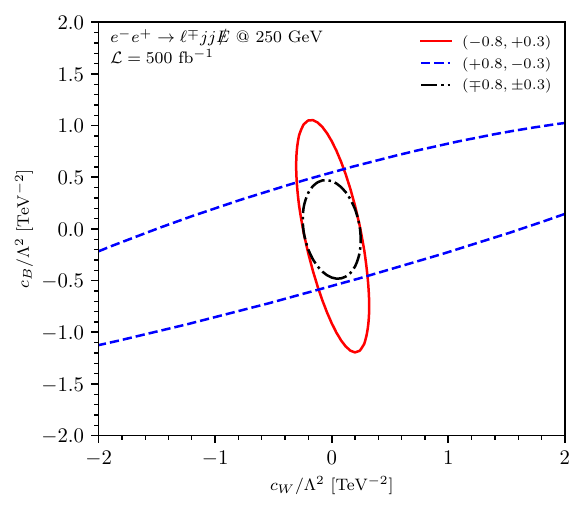}
    \includegraphics[width=0.325\textwidth]{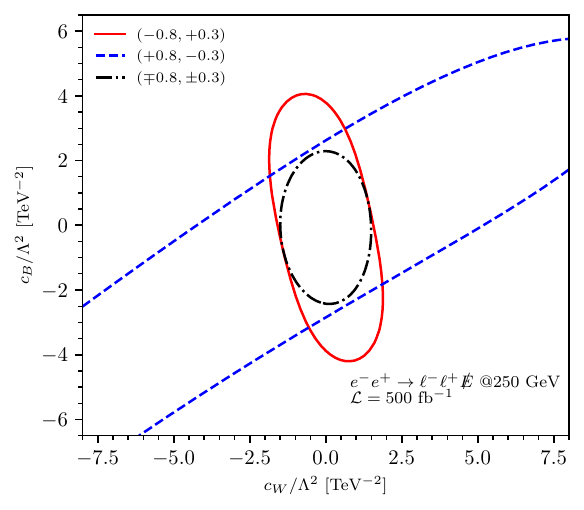}
    \includegraphics[width=0.325\textwidth]{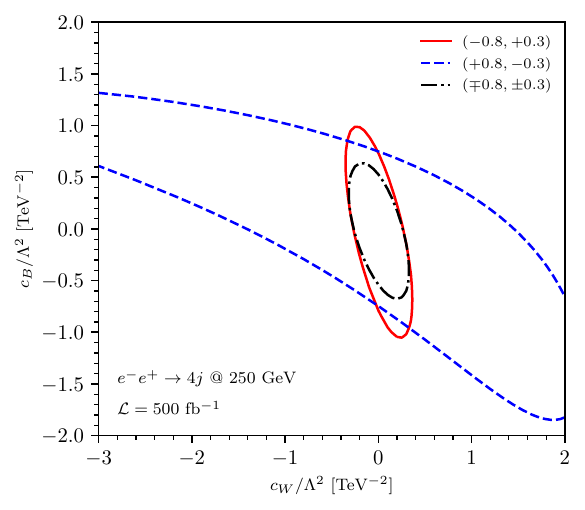}
    \caption{\label{fig:polchi}$95\%$ CL contours in a $c_W-c_B$ plane for a cross~section and asymmetries for three different final states viz. $\ell^\mp j j\slashed{e}$ (left panel), $\ell^-\ell^+\slashed{E}$ (middle panel) and $4j$ (right panel) obtained at $\sqrt{s}=250$ GeV, $\mathcal{L}=500$ fb$^{-1}$ and zero systematic errors. We obtain contours for three different initial beam polarization settings viz. $(P_{e^-},P_{e^+}) = (-0.8,+0.3)$ (solid red curve), flipped IBP (dashed blue curve) and their combination is highlighted by dashed black curve.}
\end{figure}

For a demonstration purpose, we show in Fig.~\ref{fig:polchi} the two dimensional (2D) likelihood profile in $c_{W}-c_B$ plane at $\mathcal{L}=500$ fb$^{-1}$ with zero systematics uncertainty for all three different final states. The IBP used are $(P_{e^-},P_{e^+}) = (-0.8,+0.3)$ and $(+0.8,-0.3)$. The allowed parameter space in the $c_W-c_B$ plane shows a different orientation for two set of IBP. The intersection of the two regions, which corresponds to the combined $\Delta\chi^2$ contributions, is therefore smaller than provided by different IBP alone, for all three different final states. Thus, the upcoming ILC would be a perfect experiment to utilize the potential of two sets of beam polarization to better study the deviation from the SM predictions.

\begin{figure}[!t]
    \centering
    \includegraphics[width=0.325\textwidth]{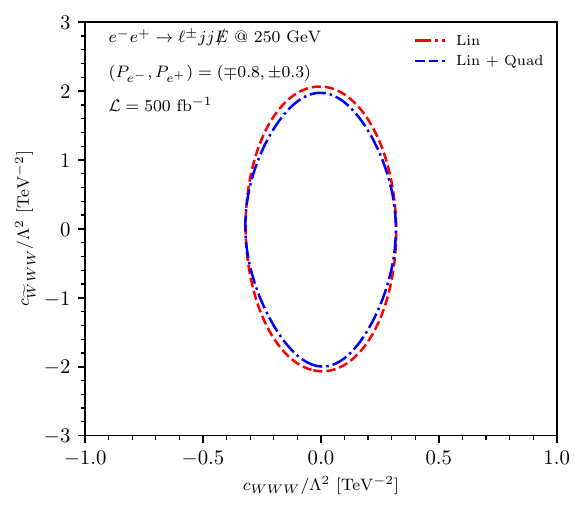}
    \includegraphics[width=0.325\textwidth]{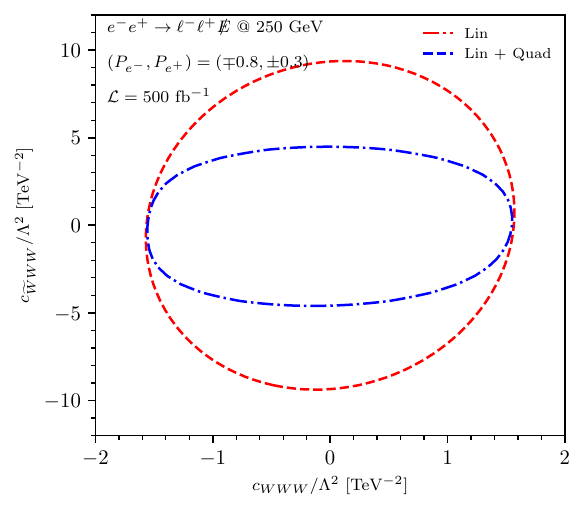}
    \includegraphics[width=0.325\textwidth]{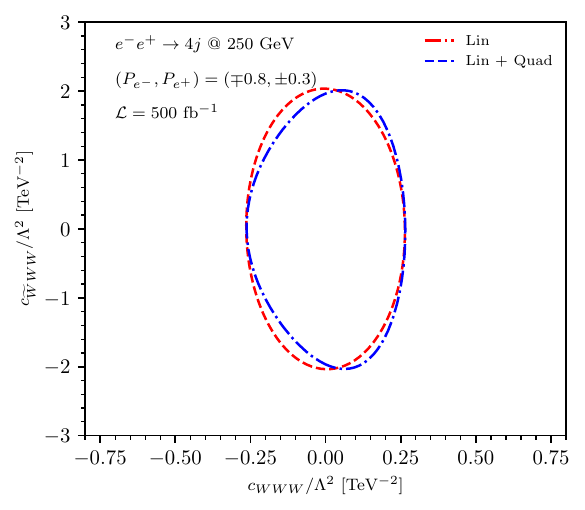}
    \caption{\label{fig:linvsquad} $95\%$ CL contours in a $c_{WWW}-c_{\widetilde{W}WW}$ plane at the level of interference (dashed red curve) and quadratic (dashed blue curve) for three different final state. The contours are obtained for $\mathcal{L}=500$ fb$^{-1}$ with $\mathcal{L}/2$ for each set of initial beam polarization. We do no consider the systematic errors in this analysis.}
\end{figure}

Next, we quantify how does the truncation of SMEFT at leading order actually plays role in the constraining of anomalous couplings. The truncation affects CP-even and CP-odd coefficients differently. The cross section and the 44 CP-even asymmetries are linear in $c_{WWW}$, $c_W$ and $c_B$, whereas $c_{\widetilde{WWW}}$ and $c_{\widetilde W}$ enter these 45 observables only at $\mathcal{O}(\Lambda^{-4})$ (Eq.~\eqref{eqn:cpeven}). The 36 CP-odd asymmetries are instead linear in the CP-odd coefficients (Eq.~\eqref{eqn:cpodd}). A fit at $\mathcal{O}(\Lambda^{-2})$ thus constrains $c_{\widetilde{WWW}}$ and $c_{\widetilde W}$ through the CP-odd asymmetries only, and the quadratic terms add the information of the CP-even observables.

Fig.~\ref{fig:linvsquad} shows the two-dimensional $95\%$ CL regions in the $c_{WWW}$--$c_{\widetilde{W}WW}$ plane for both truncations at $\mathcal{L}=500~\mathrm{fb}^{-1}$ without systematic uncertainties.  We note that in the case of CP-even coupling $c_{WWW}$, the interference terms provides a dominant contribution to the $\chi^2$ for all three final states. However, in the case of CP-odd $c_{\widetilde{W}WW}$ coupling, we note a significant contribution arising from the quadratic level.  We also note that this behaviour varies with different final states, for e.g., in the case of $\ell^-\ell^+\slashed{E}$ final state, the contour shrinks along $y$-axis by a factor of $\sim 3$ on adding a contribution from quadratic term. For the $\ell^\mp jj \slashed{E}$ final state, the contour shrink by $\sim 10\%$ along $y$-axis on adding the quadratic contribution. In the case of $4j$ final state, the impact of quadratic remains below a percent level for both axis.

\begin{figure}[!t]
    \centering
    \includegraphics[width=0.99\textwidth]{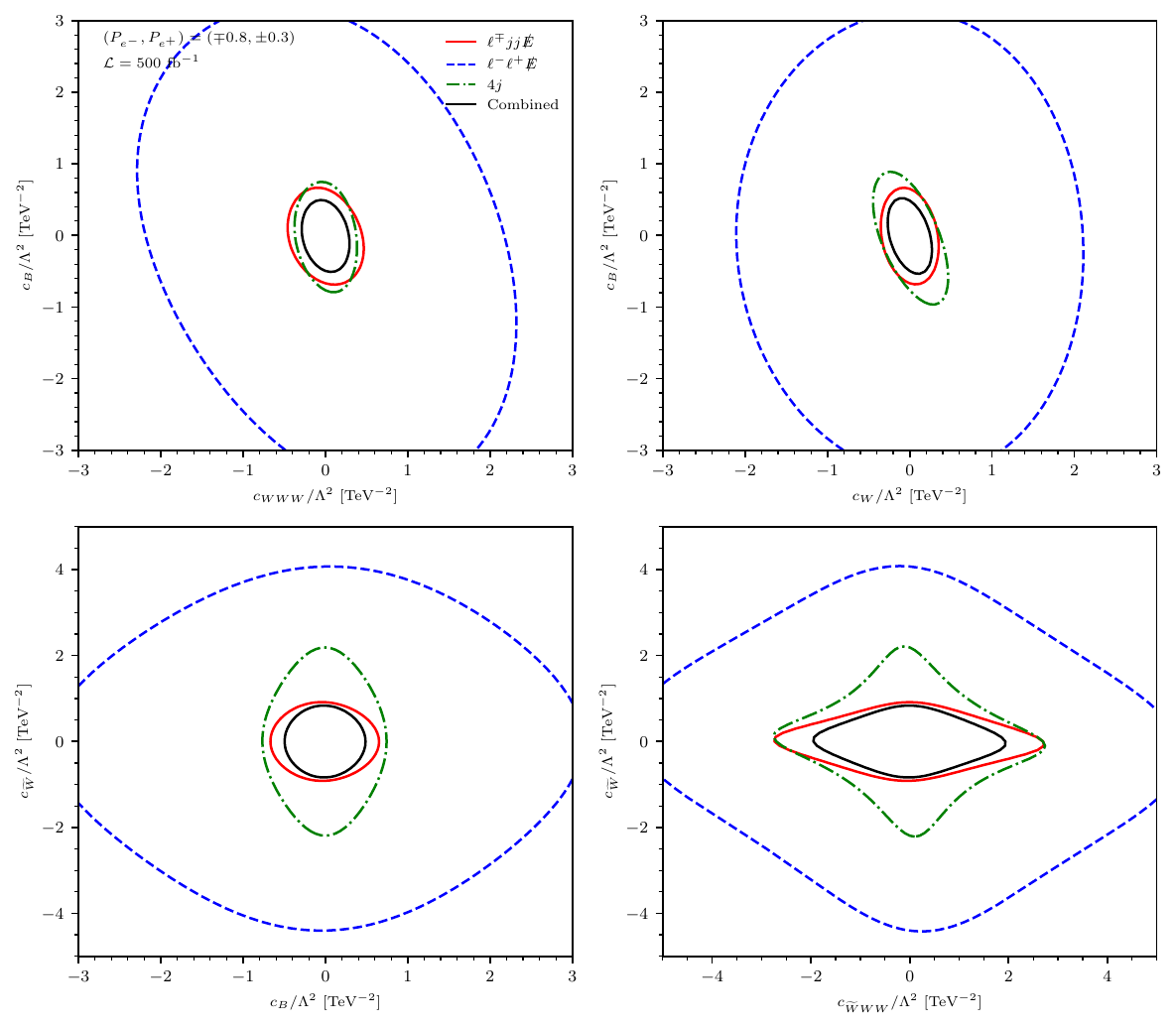}
    \caption{\label{fig:channelwise} $95\%$ CL contours for a set of asymmetries and cross~section for three different final state viz. $\ell^\mp jj \slashed{E}$ (solid red), $\ell^-\ell^+\slashed{E}$ (dashed blue) and $4j$ (dashed green) along with their combined contribution (solid black). The contours are obtained at $\mathcal{L}=500$ fb$^{-2}$ with $\mathcal{L}/2$ for each set of beam polarization. No systematic errors are considered here.}
\end{figure}

We also compare the sensitivity of each final states to the anomalous couplings. In Fig.~\ref{fig:channelwise}, we highlight four different pair of WCs viz. $c_{WWW}-c_B$, $c_W-c_B$, $c_B-c_{\widetilde{W}}$ and $c_{\widetilde{W}WW}-c_{\widetilde{W}}$ in $\mathcal{L}=500$ fb$^{-1}$ with $\mathcal{L}/2$ for each IBP with zero systematic errors. For all combinations, the full-leptonic channels present the least sensitivity. The semi-leptonic and full-hadronic provides a competitive bounds to the CP-even WCs and one CP-odd $c_{\widetilde{W}WW}$ coupling. For the $c_{\widetilde{W}}$ coupling, the dominant contribution is given by semi-leptonic channel. Moreover, in the $c_{WWW}-c_B$ plane, the semi-leptonic and full-hadronic channel offers a orthogonal cuts which leads to the overall tighter bounds.

We further investigate the dependence of the sensitivity on the integrated luminosity. We consider four benchmark luminosities motivated by the staged running scenarios commonly adopted in phenomenological studies of the ILC~\cite{ILCInternationalDevelopmentTeam:2022izu}. These span the nominal baseline $\mathcal{L}$ of \(500~\mathrm{fb}^{-1}\) at \(\sqrt{s}=250~\mathrm{GeV}\), the projected \(250~\mathrm{GeV}\) Higgs-factory programme with \(2~\mathrm{ab}^{-1}\), the \(500~\mathrm{GeV}\) stage with \(4~\mathrm{ab}^{-1}\), and a potential upgrade to \(\sqrt{s}=1~\mathrm{TeV}\) with an illustrative integrated luminosity of \(8~\mathrm{ab}^{-1}\). Accordingly, we scan over
\begin{align}
    \label{eq:lumis}
    \mathcal{L} \in \{500,\,2000,\,4000,\,8000\}~\mathrm{fb}^{-1},
\end{align}
with $\mathcal{L}/2$ for each set of IBP.

In Table~\ref{tab:onecom}, we list down the one parameter bound on five WCs at different luminosity listed in Eq.~\eqref{eq:lumis} with zero systematics. As expected, the bounds become progressively tighter with increasing integrated luminosity. Increasing \(\mathcal{L}\) from \(500\) to \(2000~\mathrm{fb}^{-1}\) improves the constraints by approximately a factor of two, while a further increase to \(8000~\mathrm{fb}^{-1}\) yields an overall improvement of roughly a factor of four across all five WCs. This behaviour is consistent with the expected $1/\sqrt{\mathcal{L}}$ dependence of statistically dominated uncertainties. 
\begin{table}[!t]
	\centering
    \caption{\label{tab:onecom}$95\%$ CL one parameter limits of Wilson coefficients obtained using cross~section and asymmetries for three different final states and also their combinations. The limits are obtained at $\sqrt{s}=250$ GeV with zero systematics for four different values of integrated luminosity given in Eq.~\eqref{eq:lumis}. For each initial beam polarization, the effective luminosity used is $\mathcal{L}/2$. }
	\renewcommand{\arraystretch}{1.3}
	\begin{tabular}{ccccc}
	\hline 
	WCs & $500$ fb$^{-1}$ & $2000$ fb$^{-1}$ & $4000$ fb$^{-1}$ & $8000$ fb$^{-1}$ \\
	\hline
	$c_{WWW}/\Lambda^2$ & $[-0.23,+0.23]$ & $[-0.11,+0.11]$ & $[-0.08,+0.08]$ & $[-0.06,+0.06]$ 
	\\
	$c_{W}/\Lambda^2$&$[-0.20,+0.20]$&$[-0.10,+0.10]$&$[-0.07,+0.07]$&$[-0.05,+0.05]$
	\\
	$c_{B}/\Lambda^2$ & $[-0.40,+0.39]$&$[-0.20,+0.20]$&$[-0.14,+0.14]$&$[-0.10,+0.10]$
	\\
	$c_{\widetilde{W}WW}/\Lambda^2$&$[-1.58,+1.57]$&$[-0.81,+0.80]$&$[-0.57,+0.57]$&$[-0.41,+0.40]$
	\\
	$c_{\widetilde{W}}/\Lambda^2$&$[-0.67,+0.67]$&$[-0.34,+0.34]$&$[-0.24,+0.24]$&$[-0.17,+0.17]$
	\\
	\hline 
	\end{tabular}
\end{table}

To benchmark our results, we compare them with the current experimental constraints and representative HL-LHC projections, as summarized in Table~\ref{tab:explim}. We begin with the baseline ILC scenario at \(\sqrt{s}=250~\mathrm{GeV}\), corresponding to an integrated luminosity of \(500~\mathrm{fb}^{-1}\). Relative to the existing experimental bounds, our limits on \(c_B/\Lambda^2\) improve by approximately a factor of \(22\), while those on \(c_{WWW}/\Lambda^2\) and \(c_W/\Lambda^2\) improve by factors of about \(4\) and \(12\), respectively, when comparing the overall widths of the allowed intervals. For the CP-odd coefficients, our constraint on \(c_{\widetilde W WW}/\Lambda^2\) is weaker than the current experimental bound, whereas the constraint on \(c_{\widetilde W}/\Lambda^2\) is substantially stronger, improving by roughly a factor of \(30\).  Compared with HL-LHC projections, our limits are stronger for CP-even couplings by factors of about $1.3-10.0$, but weaker for \(c_{\widetilde{W}WW}/\Lambda^2\) and \(c_{\widetilde{W}}/\Lambda^2\) by factors of about \(79\) and \(8.4\), respectively. Further, the momentum-dependent structure of the operators leads to enhanced sensitivity at the higher energies accessible at hadron colliders. In contrast, our analysis is performed at $\sqrt{s}=250$ GeV. Nevertheless, the use of angular asymmetries significantly improves the sensitivity, compensating for the lower energy and yielding constraints that remain competitive.

\begin{table}[!t]
        \centering
        \caption{\label{tab:explim}One parameter experimental $95\%$ CL limits and the projections at HL-LHC. }
		\renewcommand{\arraystretch}{1.3}
		\begin{tabular}{ccc}
        \hline
        WCs     & Experimental & HL-LHC Projections  \\
        \hline
        $c_{WWW}/\Lambda^2$ & $[-0.90,+0.91]$~\cite{CMS:2021foa}  & $[-0.3,+0.3]$~\cite{Eboli:2024wgy} \\
        $c_{W}/\Lambda^2$ & $[-2.50,+0.30]$~\cite{CMS:2021icx} & $[-0.8,+1.1]$~\cite{Eboli:2024wgy} \\
        $c_{B}/\Lambda^2$ & $[-8.78,+8.54]$~\cite{CMS:2019ppl} & $[-3.7,+4.0]$~\cite{Eboli:2024wgy}\\
        $c_{\widetilde{W}WW}/\Lambda^2$ & $[-0.45,+0.45]$~\cite{CMS:2021foa} & $[-0.02,+0.02]$~\cite{Biekotter:2021int} \\
        $c_{\widetilde{W}}/\Lambda^2$ & $[-20.0,+20.0]$~\cite{CMS:2021foa} & $[-0.08,+0.08]$~\cite{Biekotter:2021int} \\
        \hline
        \end{tabular}
    \end{table}
 
\subsection{MCMC analysis with combined channels}
	\label{sec:mcmc}
	To simultaneously constrain the five WCs, we perform a Markov-Chain-Monte-Carlo (MCMC) analysis based on a global likelihood function constructed from all three decay channels. Each channel is evaluated for both beam polarization configurations, $(P_{e^-}, P_{e^+}) = (\pm80\%, \mp30\%)$.
	The likelihood function is defined as
	\begin{equation}
		\mathcal{L}(\vec{C}) \propto 
		\exp\!\left[-\frac{1}{2}\Delta\chi^2_{\text{tot}}(\vec{C})\right],
	\end{equation}
	where the total chi-squared is expressed as the sum of individual contributions from each final state topology as given in Eq.~\eqref{eqn:chisq_cov}. Each $\Delta\chi^2$ term encapsulates the deviation of the predicted cross~sections and weighted asymmetries from the SM expectations, incorporating both statistical and systematic uncertainties. We choose four different combination of systematic error on cross~section and asymmetries as
\begin{align}
    \label{eq:systs}
    (\epsilon_\sigma,\epsilon_A) = \{(0,0),(0.5\%,0.1\%),(1\%,0.2\%),(2\%,0.3\%)\}.
\end{align}
The selected systematic uncertainties are intended as illustrative benchmark scenarios rather than definitive projections for the ILC. Their magnitudes are motivated by precision studies of future \(e^-e^+\) colliders, which demonstrate the potential for sub-percent control of experimental uncertainties and substantial cancellation of correlated effects in asymmetry measurements~\cite{Asner:2013psa}.


\begin{figure}[!t]
    \centering
    \includegraphics[width=0.49\textwidth]{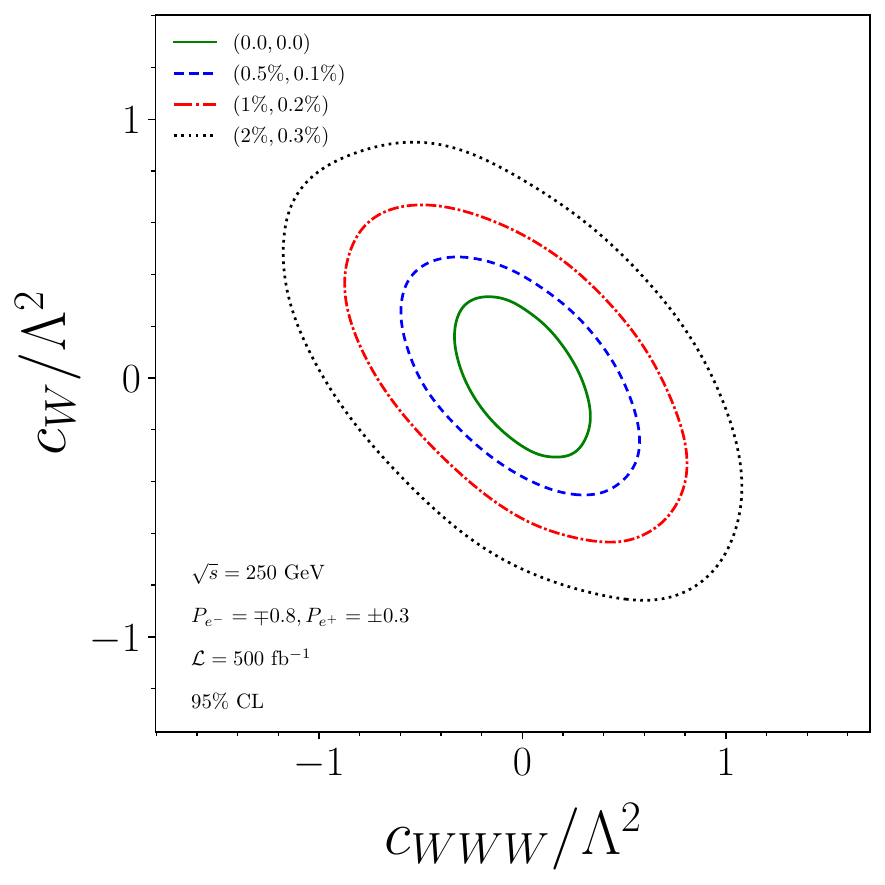}
    \includegraphics[width=0.49\textwidth]{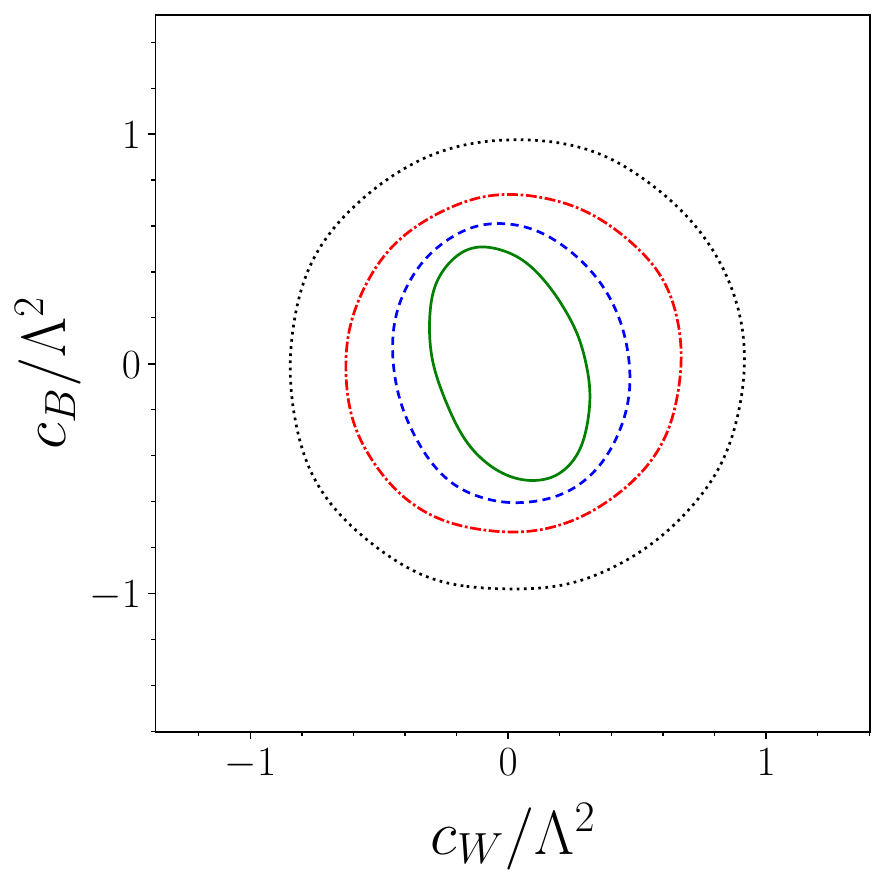}
    \includegraphics[width=0.49\textwidth]{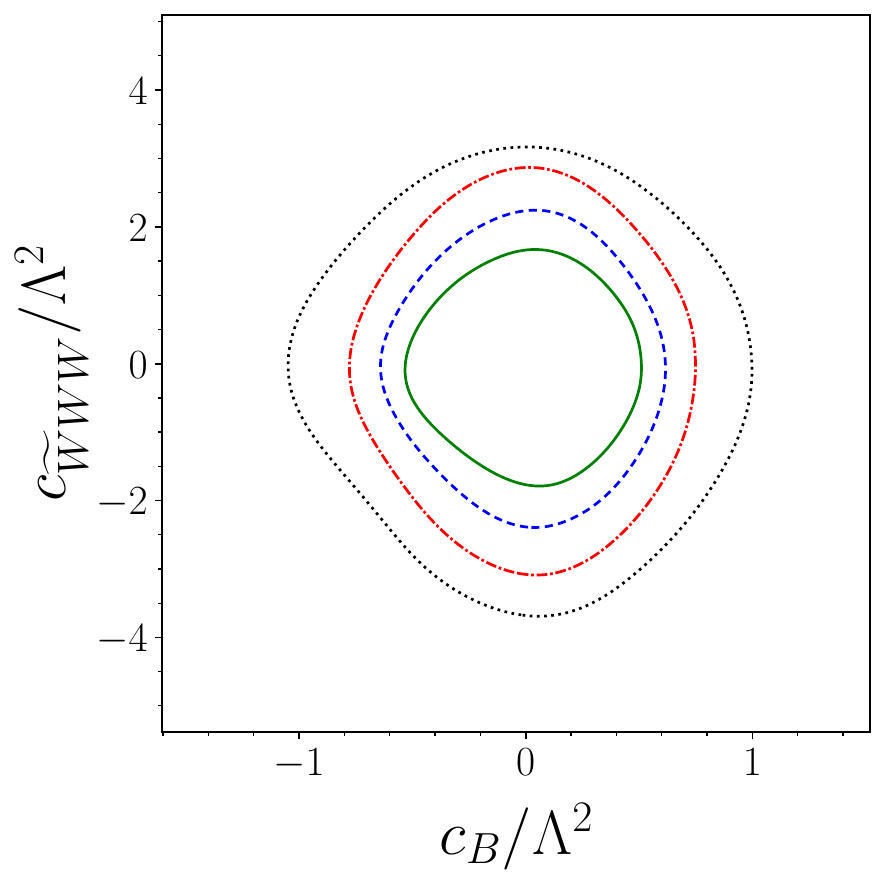}
    \includegraphics[width=0.49\textwidth]{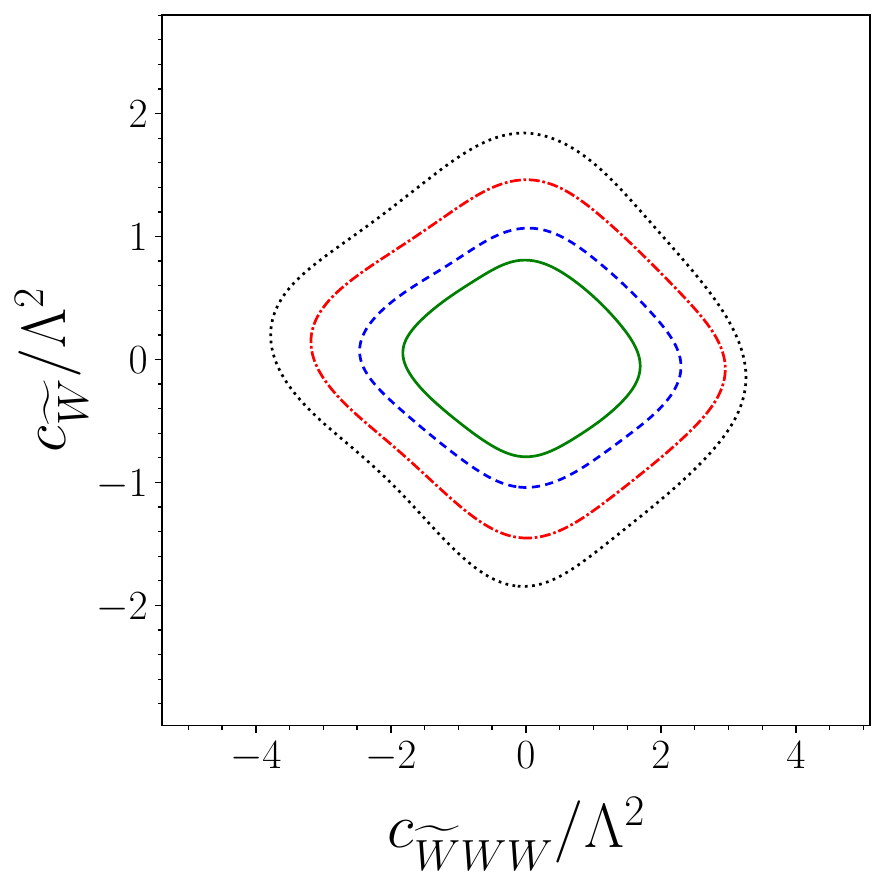}
    \caption{\label{fig:mcmcsyst}Marginalized 2-D projection at $95\%$ CL from MCMC for a different values of systematic errors. The projections are obtained by combining all three final states at $\mathcal{L}=500$ fb$^{-1}$.}
\end{figure}

The sampling of the posterior distribution is performed using a random-walk Metropolis algorithm~\cite{Metropolis:1953am,Hastings:1970aa} in the five-dimensional parameter space of the WCs, $\vec{c}$. At each iteration, a trial point $\vec{c}'$ is generated by applying Gaussian perturbations to the current state $\vec{c}$, and the move is accepted with probability
	\begin{equation}
		\rho = \min\!\left[1,\, 
		\exp\!\left(-\frac{1}{2}\left[\Delta\chi^2_{\text{tot}}(\vec{C}') - 
		\Delta\chi^2_{\text{tot}}(\vec{C})\right]\right)\right].
	\end{equation}
	This procedure efficiently explores the correlated parameter space and yields posterior probability distributions for the WCs that incorporate the combined sensitivity from all decay channels and beam polarizations. The one and two-dimensional densities are computed from the generated MCMC samples using Kernel Density Evaluation implemented in \textsc{ GetDist}~\cite{Lewis:2019xzd}.

    	Figure~\ref{fig:mcmcsyst} shows the $95\%$~CL marginalized two-parameter posterior projections for fixed integrated luminosity $\mathcal{L}=500~\mathrm{fb}^{-1}$ under different assumptions on systematic uncertainties. A clear contraction of the allowed regions is observed when going from the conservative scenario $(\epsilon_\sigma,\epsilon_A)=(2\%,\,0.3\%)$ to the idealistic case. This reduction is visible across all highlighted operator planes, with the $(c_{WWW},\,c_{W})$ projection exhibiting a stronger negative correlation, while the other three planes remain largely symmetric, indicating weak correlations.

    Quantitatively, the transition from $(2\%,\,0.3\%)$ to $(1\%,\,0.2\%)$ reduces the allowed parameter space by a factor of $\sim 1.15$--$1.38$, depending on the operator combination. On further reducing the errors to the idealized scenario, the contours shrink by a factor of $\sim 1.4-2.5$, with minimal shrinkage along $c_B$ axis. This indicates that, at $\mathcal{L}=500~\mathrm{fb}^{-1}$, the sensitivity can be further increased on reducing the overall systematic errors on rate and asymmetries, which is one of the promising goal of future Higgs-machine.

    We also performed a MCMC analysis for different values of integrated luminosity as given in Eq.~\eqref{eq:lumis} and the corresponding $95\%$ CL marginalized limits on all five WCs are listed in Table~\ref{Tab:final}. The results in Table~\ref{Tab:final} illustrate the complementary roles of integrated luminosity and systematic control in constraining the five WCs simultaneously. In the idealized zero-systematics scenario, increasing the luminosity from \(500\) to \(8000~\mathrm{fb}^{-1}\) improves the constraints by approximately a factor of four, consistent with the expected \(\mathcal{L}^{-1/2}\) scaling in the statistics-dominated regime. Once systematic uncertainties are included, however, the improvement becomes substantially more modest, particularly for the more conservative uncertainty assumptions. This behaviour indicates that increasing the data sample alone is insufficient to fully exploit the potential precision of the collider: systematic control becomes increasingly important as the statistical uncertainties decrease. The results therefore highlight the value of combining high integrated luminosity with precise measurements of cross sections and asymmetries. In this respect, the ILC offers a promising environment for simultaneous constraints on anomalous triple gauge couplings, with the potential to probe complementary directions in the multidimensional WC parameter space.
 \begin{table}[!htb]
	\centering
	\renewcommand{\arraystretch}{1.5} 
	\caption{\label{Tab:final} Marginalized 95$\%$ CL limits on the dimension-6 operators obtained with different values of integrated luminosity and systematic error as given in Eq.~(\ref{eq:systs}).}
    \scriptsize
	\begin{tabular*}{1\textwidth}{@{\extracolsep{\fill}}ccccccc@{}}\hline
		$\mathcal{L}$ (fb$^{-1}$)&($\epsilon_\sigma,\epsilon_{\mathcal{A}}$)& $c_{WWW}/\Lambda^2$ & $c_W/\Lambda^2$ &$c_B/\Lambda^2$&$c_{\widetilde{W}WW}/\Lambda^2$ & $c_{\widetilde{W}}/\Lambda^2$ \\ \hline			
			  & $(0,0)$&$[-0.27,+0.27]$&$[-0.24,+0.25]$&$[-0.41,+0.41]$&$[-1.41,+1.30]$&$[-0.60,+0.62]$\\
			  & $(0.005,0.001)$&$[-0.48,+0.47]$ &$[-0.36,+0.38]$&$[-0.49,+0.49]$&$[-1.89,+1.75]$&$[-0.79,+0.81]$\\
	    $500$ & $(0.01,0.002)$&$[-0.70,+0.64]$ &$[-0.50,+0.54]$&$[-0.59,+0.59]$&$[-2.45,+2.20]$&$[-1.09,+1.09]$\\
	    	  & $(0.02,0.003)$&$[-0.95,+0.85]$ &$[-0.68,+0.74]$&$[-0.80,+0.78]$&$[-2.94,+2.48]$&$[-1.38,+1.38]$\\
		\hline
			  & $(0,0)$&$[-0.14,+0.13]$ & $[-0.12,+0.13]$ &$[-0.21,0.21]$ &$[-0.76,+0.74]$ &$[-0.32,+0.33]$\\
	    	  & $(0.005,0.001)$& $[-0.35,+0.34]$ &$[-0.26,+0.27]$ & $[-0.31,+0.31]$ &$[-1.49,+1.43]$ & $[-0.58,+0.60]$ \\
	    $2000$ & $(0.01,0.002)$& $[-0.60,+0.55]$&$[-0.42,+0.45]$ &$[-0.44,+0.44]$ &$[-2.20,+2.21]$ &$[-0.92,+0.92]$ \\
	          & $(0.02,0.003)$& $[-0.86,+0.78]$& $[-0.61,+0.65]$&$[-0.68,+0.67]$ &$[-2.79,+2.43]$ &$[-1.25,+1.26]$ \\
		\hline
			  & $(0,0)$ & $[-0.96,+0.94]$ &$[-0.88,+0.89]$ &$[-0.15,+0.15]$ &$[-0.57,+0.55]$ &$[-0.23,+0.23]$\\
			  & $(0.005,0.001)$&$[-0.32,+0.31]$&$[-0.24,+0.24]$ &$[-0.26,+0.26]$ &$[-1.39,+1.33]$ &$[-0.52,+0.54]$ \\
		$4000$& $(0.01,0.002)$& $[-0.57,+0.52]$&$[-0.40,+0.42]$ &$[-0.40,+0.40]$ &$[-2.15,+1.99]$ &$[-0.88,+0.90]$ \\
			  & $(0.02,0.003)$& $[-0.83,-0.77]$& $[-0.59,+0.63]$&$[-0.65,+0.64]$ &$[-2.76,+2.40]$ &$[-1.21,+1.23]$ \\
		\hline
			  & $(0,0)$&$[-0.07,+0.07]$ &$[-0.06,+0.06]$ &$[-0.11,+0.11]$ &$[-0.40,+0.40]$ &$[-0.17,+0.17]$\\
			  & $(0.005,0.001)$&$[-0.30,+0.29]$&$[-0.22,+0.22]$ &$[-0.23,+0.22]$ &$[-1.32,+1.29]$ &$[-0.49,+0.50]$ \\
		$8000$ & $(0.01,0.002)$&$[-0.55,+0.51]$ &$[-0.39,+0.40]$ &$[-0.37,+0.38]$ &$[-2.11,+1.94]$ &$[-0.86,+0.87]$ \\
		& $(0.02,0.003)$& $[-0.83,+0.75]$& $[-0.58,+0.62]$&$[-0.63,+0.62]$ &$[-2.74,+2.37]$ &$[-1.21,+1.22]$ \\
		\hline
	\end{tabular*}
\end{table}
\section{Conclusion}
\label{sec:concl}

We have performed a leading-order study of charged triple gauge couplings
in $e^-e^+\to W^-W^+$ at $\sqrt{s}=250$~GeV with longitudinally polarized
beams, within the SMEFT framework and the HISZ basis. The five
dimension-six operators $\mathcal{O}_{WWW}$, $\mathcal{O}_{W}$,
$\mathcal{O}_{B}$, $\mathcal{O}_{\widetilde{W}WW}$ and
$\mathcal{O}_{\widetilde{W}}$ were constrained using the cross section and
the full set of 80 polarization and spin-correlation asymmetries of the
$W^-W^+$ system. All three decay channels viz. $\ell^\pm jj\slashed{E}$, $4j$ and $\ell^-\ell^+\slashed{E}$, were simulated using \textsc{MadGraph5\_aMC@NLO}, showered and hadronized with
\textsc{Pythia8} and detector effect implemented via. \textsc{Delphes}. All leading-order amplitudes contributing to each final state were retained, so that
non-resonant topologies and their interference with $W^-W^+$ production
are included.

The event selection and reconstruction were tailored to each channel. In
the semi-leptonic channel the signal was selected with cuts on
$m_{j_1j_2}$ and $m_T^{W}$, while the neutrino momentum was fixed by the on-shell $W$ mass constraint. The flavor tagging of the two hardest light jets in two class i.e., \emph{up}-type and \emph{down}-type were performed using BDT with an accuracy of approximately $81\%$ (AUC $=0.90$). In the fully hadronic channel a multi-class BDT separated $W^-W^+$ from $ZZ$ and
QCD four-jet production and raised the signal purity from approximately
$86\%$ to $97\%$. The $W^\pm$ candidates were then assigned through the
$p_T$-weighted jet charge, with a tagging efficiency of approximately $0.75$
for $\kappa=0.2$. In the fully leptonic channel the signal was isolated with cuts on $p_T^{\ell^-}$ and $m_{\ell^-\ell^+}$, and the two neutrino momenta were
reconstructed analytically from the on-shell and beam-energy constraints,
with the finite $W$ width accounted for by a scan over trial masses. The
reconstructed \textsc{Delphes} level transverse and longitudinal neutrino momenta show Pearson correlation coefficients with the parton-level values of $0.80$ to $0.82$ and $0.70$ to $0.71$, respectively.

The two different set of initial beam polarization offers a orthogonal constraint on the parameter space translating to overall tighter allowed space.
Different channels constrain different directions in the space of Wilson
coefficients. The semi-leptonic and fully hadronic channels give competitive
bounds on the CP-even coefficients and on $c_{\widetilde{W}WW}$. In the
$c_{WWW}$--$c_B$ plane their allowed regions are oriented differently, so
that their combination is tighter than either alone. The constraint on
$c_{\widetilde{W}}$ is dominated by the semi-leptonic channel. The fully
leptonic channel is the least sensitive in all planes examined, owing to its
small rate, and it improves the combination only marginally. The quadratic term ($\mathcal{O}(\Lambda^{-4})$)in the dimension-six amplitude matter mainly for the CP-odd $c_{\widetilde{W}WW}$ in the fully leptonic channel,
where truncating at $\mathcal{O}(\Lambda^{-2})$ enlarges the allowed region
by a factor of approximately three. The effect is about $10\%$ in the semi-leptonic channel and below $1\%$ in the fully hadronic one.

For the baseline luminosity of $500~\mathrm{fb}^{-1}$ with $\mathcal{L}/2$ for each beam polarization configurations $(P_{e^-},P_{e^+})=(\mp0.8,\pm0.3)$,
and in the absence of systematic uncertainties, the combined one-parameter
bounds at $95\%$ CL are
$c_{WWW}\in[-0.23,0.23]$, $c_{W}\in[-0.20,0.20]$,
$c_{B}\in[-0.40,0.39]$, $c_{\widetilde{W}WW}\in[-1.58,1.57]$
and $c_{\widetilde{W}}\in[-0.67,0.67]$ for $\Lambda=1$ TeV. Compared to the existing experimental bounds, our limits on $c_B$ improve by approximately a factor of $22$, while those on $c_{WWW}$ and $c_W$ improve by factors of about $4$ and $12$, respectively, when comparing the overall widths of the allowed intervals. For the CP-odd coefficients, our constraint on $c_{\widetilde W WW}$ is weaker than the current experimental bound, whereas the constraint on $c_{\widetilde W}$ improves by roughly a factor of $30$. Further, in comparison to the HL-LHC projections, our limits are stronger for CP-even couplings by factors of about $1.3-10.0$, but weaker for \(c_{\widetilde{W}WW}\) and \(c_{\widetilde{W}}\) by factors of about \(79\) and \(8.4\), respectively. The momentum-dependent structure of the operators leads to enhanced sensitivity at the higher energies accessible at hadron colliders. In contrast, our analysis is performed at $\sqrt{s}=250$ GeV. Nevertheless, the use of angular asymmetries significantly improves the sensitivity, compensating for the lower energy and yielding constraints that remain competitive.

The dependence on integrated luminosity and on systematic uncertainties was studied for $\mathcal{L}\in\{0.5,2,4,8\}~\mathrm{ab}^{-1}$ and four benchmarks $(\epsilon_\sigma,\epsilon_A)=(0,0)$, $(0.5\%,0.1\%)$, $(1\%,0.2\%)$ and
$(2\%,0.3\%)$. Without systematic uncertainties the limits improve as $1/\sqrt{\mathcal{L}}$, by approximately a factor of four from $0.5$ to $8~\mathrm{ab}^{-1}$. Once systematic uncertainties are included, they limit
the sensitivity already at $500~\mathrm{fb}^{-1}$: for $(\epsilon_\sigma,\epsilon_A)=(2\%,0.3\%)$ the marginalized bound on $c_{WWW}$ improves by only about $12\%$ from increasing $\mathcal{L}$ from $0.5$ and $8~\mathrm{ab}^{-1}$. Improving the constraints on all five coefficients therefore requires control of the systematic uncertainties on the rates and asymmetries.
	
	\acknowledgments
	Authors thanks R. Rahaman for discussion during the initial phase of this project. 
\bibliographystyle{JHEP} 
\bibliography{biblio.bib}

\end{document}